\documentclass[iop,twocolumn,twocolappendix]{emulateapj}
\usepackage{color,natbib}

\shortauthors{Kuntz et al}
\shorttitle{Observations on SWCX}

\slugcomment{Version 1, KDK, \today}
\newcommand{\rosat}{{\it ROSAT}}

\newcommand{\chandra}{{\it Chandra}}
\newcommand{\xmm}{{\it XMM-Newton}}
\newcommand{\suzaku}{{\it Suzaku}}
\newcommand{\ace}{{\it ACE}}
\newcommand{\imp}{{\it IMP 8}}

\newcommand\oqkev{$\frac{1}{4}$ keV}
\newcommand\tqkev{$\frac{3}{4}$ keV}
\def\flux{\:\rm{\:cm^{-2}\:s^{-1}}}
\newcommand\re{R$_E$}

\newcommand{\bats}{{BATS-R-US}}

\begin{document}

\title{The Solar Wind Charge-Exchange Production Factor for Hydrogen}

\author{
K. D. Kuntz\altaffilmark{1}, 
Y. M. Collado-Vega\altaffilmark{2},
M. R. Collier\altaffilmark{2},
H. K. Connor\altaffilmark{2},
T. E. Cravens\altaffilmark{3},
D. Koutroumpa\altaffilmark{4},\\
F. S. Porter\altaffilmark{2},
I. P. Robertson\altaffilmark{3},
D. G. Sibeck\altaffilmark{2},
S. L. Snowden\altaffilmark{2},
N. E. Thomas\altaffilmark{2} \&
B. M. Walsh\altaffilmark{5} }
\altaffiltext{1}{The Henry A. Rowland Department of Physics and Astronomy,
Johns Hopkins University, 3400 N. Charles Street, Baltimore, MD 21218, USA;
kuntz@pha.jhu.edu}
\altaffiltext{2}{NASA/Goddard Space Flight Center,
Greenbelt, MD 20771, USA}
\altaffiltext{3}{Department of Physics and Astronomy,
University of Kansas, Lawrence, KS 66045, USA}
\altaffiltext{4}{LATMOS/IPSL/CNRS,
Universit\'{e} de Versailles Saint Quentin en Yvelines,
Guyancourt, France}
\altaffiltext{5}{Space Sciences Laboratory, University of California, Berkeley, CA 94720, USA }

\keywords{X-rays: diffuse emission}


\begin{abstract}
The production factor, or broad band averaged cross-section, for solar wind charge-exchange with hydrogen producing emission in the \rosat\ \oqkev\ (R12) band is $(3.8\pm0.2) \times 10^{-20}$ count degree$^{-2}$ cm$^4$. This value is derived from a comparison of the Long-Term (background) Enhancements in the \rosat\ All-Sky Survey with magnetohysdrodynamic simulations of the magnetosheath. This value is 1.8 to 4.5 times higher than values derived from limited atomic data, suggesting that those values may be missing a large number of faint lines. This production factor is important for deriving the exact amount of \oqkev\ band flux that is due to the Local Hot Bubble, for planning future observations in the \oqkev\ band, and for evaluating proposals for remote sensing of the magnetosheath. The same method cannot be applied to the \tqkev\ band as that band, being composed primarily of the oxygen lines, is far more sensitive to the detailed abundances and ionization balance in the solar wind. We also show, incidentally, that recent efforts to correlate \xmm\ observing geometry with magnetosheath solar wind charge-exchange emission in the oxygen lines have been, quite literally, misguided. Simulations of the inner heliosphere show that broader efforts to correlate heliospheric solar wind charge-exchange with local solar wind parameters are unlikely to produce useful results.
\end{abstract}

\section{Introduction}


\subsection{Motivation}

Solar wind charge-exchange (SWCX) occurs when a high charge state ion in the solar wind charge-exchanges with a neutral; the resultant ion is in an excited state and in the transition to the ground state a photon is produced in either the soft X-ray or extreme ultraviolet. Since the solar wind is temporally variable in density, speed, elemental abundance, and ionization fraction, the observed SWCX emission varies as a function of time, look direction, and observatory location. The charge-exchange spectrum has the same transitions, from the same ions, with nearly the same strengths, as a purely recombining plasma, and thus makes a strong contribution to the lines used for astrophysical plasma diagnostics \citep[See, for example, the discussion in ][]{wbb2008}. Because the SWCX emission will be smooth over the FOV of any recent, current, or near future X-ray instrument, and since it has no unique spectral signatures, it is a particularly problematic foreground to remove from observations of any diffuse emission that also fills the FOV, such as that expected from the Local Hot Bubble, the Galactic bulge and halo, and the Warm-Hot Intergalactic Medium. It is also problematic for many observations of clusters of galaxies, nearby galaxies, local supernova remnants, and super bubbles. 

Indeed, the uncertainty in the strength of the SWCX emission in the \rosat\ \oqkev\ band has led to significant controversy over the existence of the Local Hot Bubble (LHB). The clearest evidence for the soft X-ray emission from the LHB is significant emission in front of nearby ($\sim60$ pc) molecular clouds \citep[e.g.,][]{smv1993}. The distribution of hot gas was derived from an all-sky anti-correlation analysis \citep{sea_lhb} which showed a rather hourglass shaped distribution where the waist is in the Galactic plane, and the bulges extend towards the Galactic poles. If, however, the bulk of the emission attributed to the LHB were actually due to SWCX, then the LHB might not exist, and the distribution of the remaining hot gas could be explained in other ways \citep{wea1999}. Various other data, such as the pressure difference between the hot gas and the local interstellar clouds \citep{jenkins2009}, the distribution of \ion{O}{7}\footnote{We will use \ion{O}{7} to refer to the emission line, while we will use O$^{+6}$ to refer to the parent ion.} emission from the LHB-cool cloud interfaces \citep[][among others]{wl2008}, the shape of the local cavity \citep{sfeir1999}, and the magnetic field in the wall of the LHB \citep{ap2006} were variously deployed to argue both for and against the existence of the LHB. While it was generally recognized that some part of the soft X-ray emission previously attributed to the LHB is indeed due to SWCX, estimates of the fraction of the emission due to SWCX varied greatly \citep[compare, for example][]{dk2009,ir2009}. Further confusion arose because the more reliable calculations for the SWCX emission were made in the \tqkev\ band, which is dominated by a few strong lines, while the bulk of the LHB emission is in the \oqkev\ band where SWCX calculations are both difficult and highly uncertain. That uncertainty arises from the lack of reliable charge-exchange cross-sections for the multitude of faint lines that form the bulk of the \oqkev\ band emission.

In the intervening years, observational verification of calculations of SWCX emission \citep[e.g.,][]{dk2012} has focussed on the \ion{O}{7} and \ion{O}{8} lines because those lines are accessible to \chandra , \xmm , and \suzaku , while the \oqkev\ band is not. While such work has explored issues concerning the distribution of the neutral material with which the solar wind charge-exchanges and the structure of the solar wind itself, progress on the validation of our calculation of the SWCX emission in the \oqkev\ band has been minimal. While the number of measured or calculated charge-exchange cross-sections has grown \citep[][and several other studies in preparation]{ebit2014}, these do not yet form a sufficient database with which to calculate reliably the SWCX emission in the \oqkev\ band. However, even if all of the contributing charge-exchange cross sections and branching ratios were well known, modeling of the SWCX emission would still be problematic because of the lack of sufficiently detailed abundance and ionization state information for the solar wind.

The DXL sounding rocket observation \citep{dxl2014} of the He focussing cone (the region where the Sun gravitationally focuses the interstellar He flowing through the solar system) in the Wisconsin C-band allows a more empirical approach; the ratio of the observed X-ray emission to the model emission measure provides a broad-band abundance-weighted cross-section, or production factor. Although the Wisconsin C-band is not exactly the \rosat\ \oqkev\ band, they are sufficiently similar that scaling from one to the other does not introduce significant uncertainties. However, DXL only provides the production factor for SWCX with He (the IS H being strongly depleted near the He focussing cone), whereas a comparable emission is expected from SWCX with H, both from the interstellar H flowing through the solar system and the exospheric H in the Earth's magnetosheath. Here we derive a \rosat\ \oqkev\ band production factor for SWCX with H that requires no more than existing data and the application of relatively off-the-shelf magnetohydrodynamic (MHD) simulations. 

This work has implications beyond the LHB. Study of the Earth's magnetosheath has been carried out through {\it in situ} measurements of the particle populations and magnetic fields. However, there has been no means of imaging the entire magnetosheath at the timescales necessary for capturing the driving physical processes\footnote{Imaging of the magnetosheath with energetic neutral atoms (ENA) is possible, but only with exposures of 0.5 to 2 days \citep{petrinec2011,fuselier2010} See also the discussion by \citet{connor2014}.}. Imaging of the magnetosphere in the soft X-rays produced by SWCX with either a wide-field imager in Earth orbit, or a more traditional instrument in solar orbit can cover the entire magnetosphere \citep{storm2012}. The primary issue with such an imager is the effective cadence, which can be determined only in the light of the expected flux. Given that the magnetosheath SWCX is due primarily to charge-exchange with H, and given that that emission is primarily in the \oqkev\ band, then the production factor derived in this paper is essential for such calculation. 

\subsection{Terminology}

\begin{figure*}
\center{\includegraphics[width=8.5cm,angle=0.0]{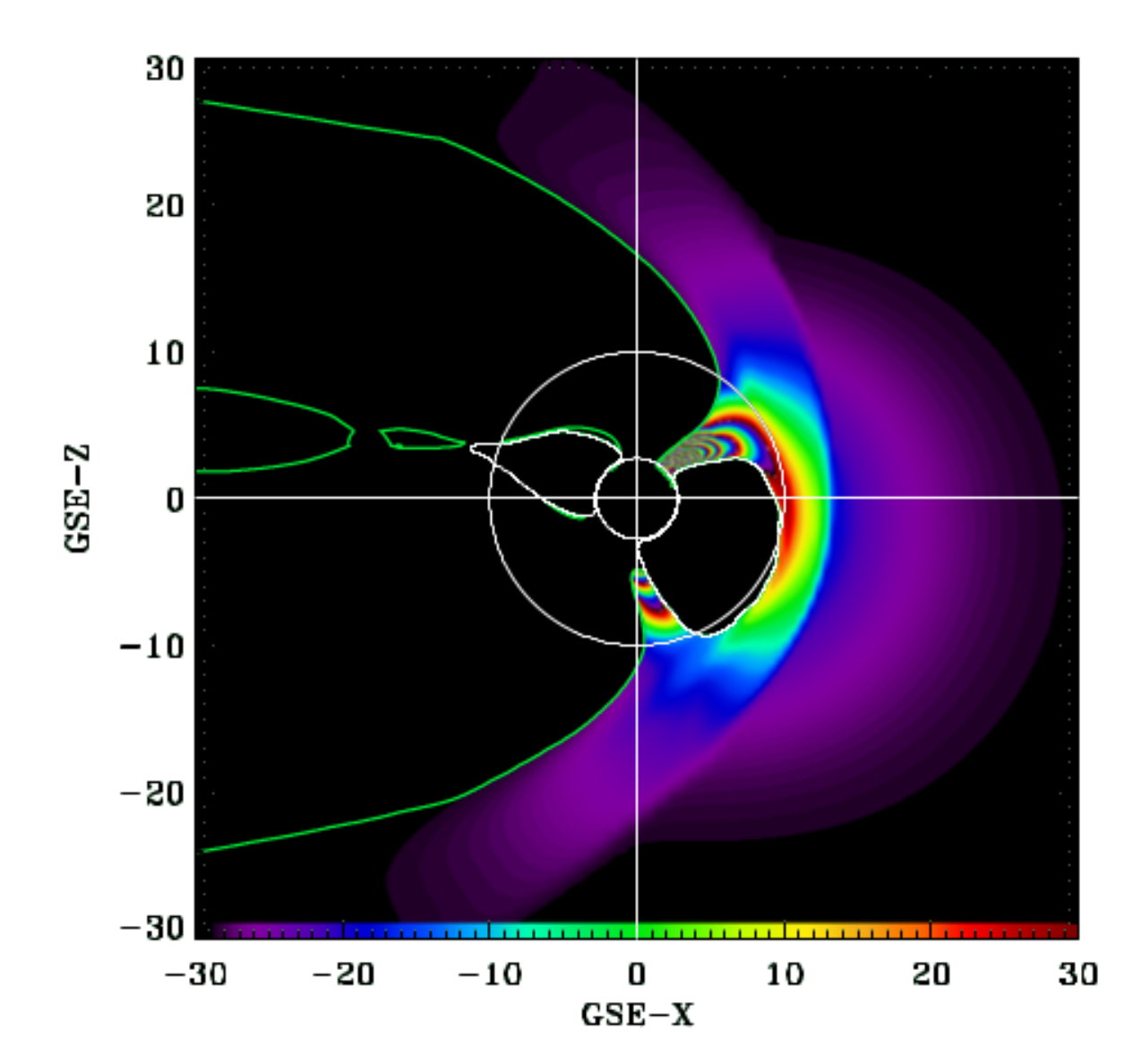}\includegraphics[width=8.5cm,angle=0.0]{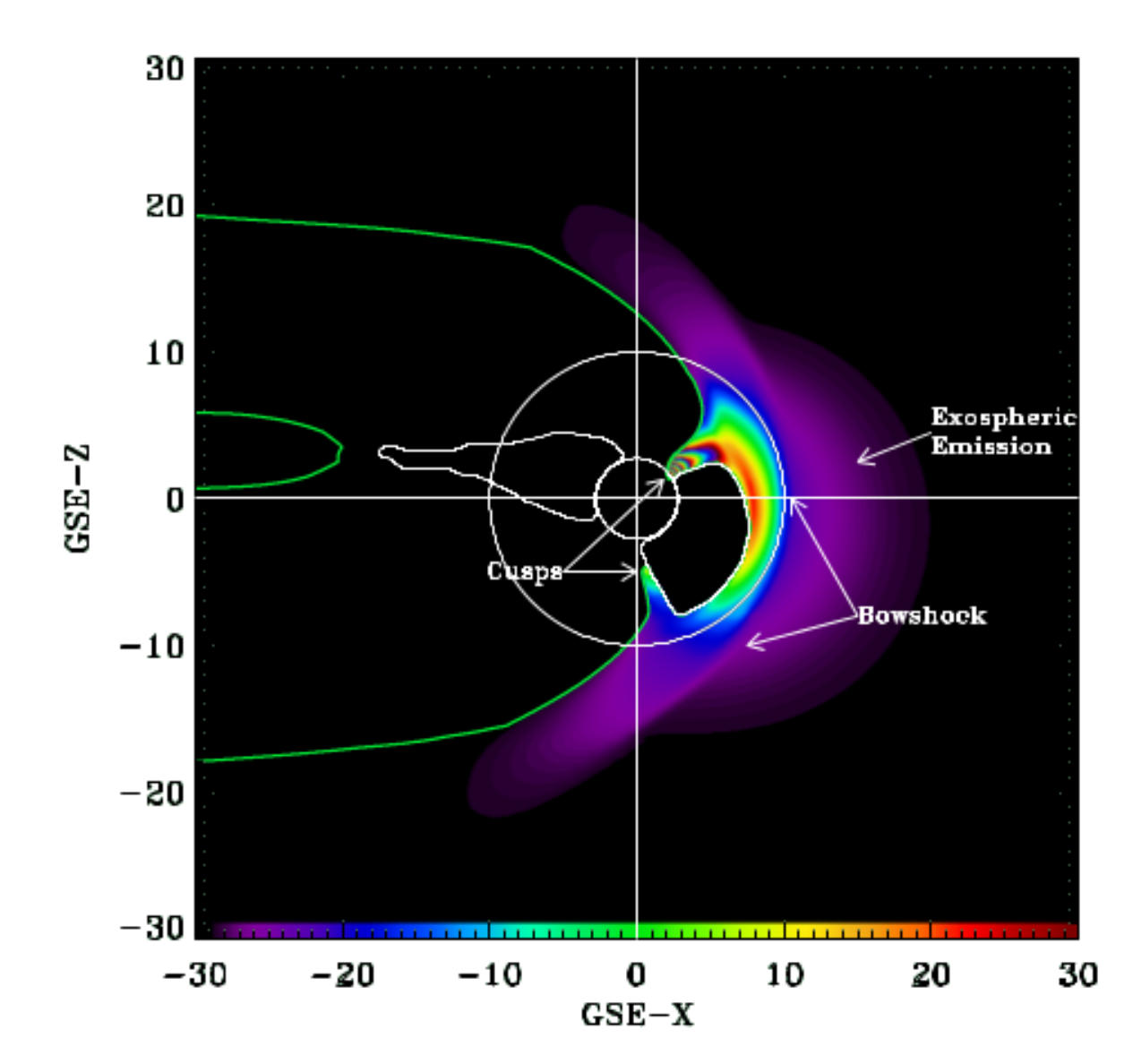}}
\caption{The magnetosphere at two different solar wind fluxes. Both images are cuts in the GSE-XZ plane. The color scale shows the relative soft X-ray emissivity as a function of position. Due to the large dynamic range, the color stretch is problematic, and so has been allowed to wrap. The {\it solid white contour} traces the region where the field lines are closed, while the {\it large white circle} marks a radius of 10 R$_E$ from the center of the Earth while the {\it small white circle} marks a radius of 2.5 R$_E$. The {\it solid green contour} traces the region removed in the course of removing the plasma not related to the solar wind. In both images the interplanetary magnetic field has $B_z=-5$ nT (southward pointing) and the time is near the summer solstice. In these images one can see the location of the magnetopause and the bowshock, as well as emission due to the interaction of the free-flowing solar wind with the exosphere. {\bf Left: }  $n_{swp}v_{swp}=2.4\times10^8$, a typical solar wind flux. {\bf Right: } $n_{swp}v_{swp}=7.8\times10^8$, a strong solar wind flux. The color bar for the right panel extends an order of magnitude higher than the color bar for the left panel. 
\label{fig:ms_demo}}
\end{figure*}

Since there are multiple sites of SWCX, besides comets and the atmospheres and surfaces of other planets, it is useful to define three different types of SWCX emission based on location and phenomenology.

The {\it Magnetosheath\footnote{The term magnetospheric SWCX has also been used (even by this author) but is inaccurate. Further, even among heliosphericists, there is some ambiguity in the meaning and application of ``magnetosphere'', so readers should be cautious.} SWCX} is due to the solar wind plasma interacting with the neutrals, primarily H, in the Earth's extended atmosphere, or exosphere. The terrestrial magnetic field is an obstacle to the charged particles that form the solar wind. The {\it magnetopause} is the surface where the ram pressure of the solar wind is balanced by the pressure of the terrestrial magnetic field. Since the solar wind is supersonic and super-Alfvenic, a roughly parabolic bow shock forms upstream of the magnetopause. The {\it magnetosheath}, the region between the magnetopause and the bow shock, contains shocked solar wind plasma. Behind the nose of the bow shock, that plasma has a density roughly four times higher than that in the free-flowing solar wind; as the plasma flows away from the nose into the flanks of the magnetosheath, its density decreases and its velocity increases. The typical {\it stand-off} distance of the magnetopause in the direction of the Sun is $\sim10$ R$_E$ from the Earth, but varies with the solar wind ram pressure as $(nv^2)^{-(\frac{1}{6})}$. The magnetosheath has a thickness of a few R$_E$ that varies with solar wind parameters. As the solar wind flux increases, the magnetopause is pushed deeper into the exosphere, so that the increased solar wind flux interacts with a higher neutral density, and the X-ray emissivity increases drastically. The continuous compression and expansion of the magnetosphere causes the magnetosheath SWCX to be strongly time-variable. Because the magnetosheath plasma fluxes decrease with distance from the nose of the bowshock, the observed magnetosheath SWCX emission is also strongly dependent on the observation geometry. Simulations of the soft X-ray emissivity in the GSE-XZ\footnote{All coordinates for the magnetosheath and related regions will be given in geocentric solar ecliptic (GSE) coordinates. In this right-handed coordinate system, the X-axis is the Earth-Sun line, the Z-axis is the ecliptic pole, and the Y-axis is defined by the right-handedness. The positive Y-axis is opposite to the direction of the Earth's motion around the Sun. See \citet{hapgood} for further explication.} plane are shown in Figure~\ref{fig:ms_demo}.

\begin{figure*}
\center{\includegraphics[width=16.0cm,angle=0.0]{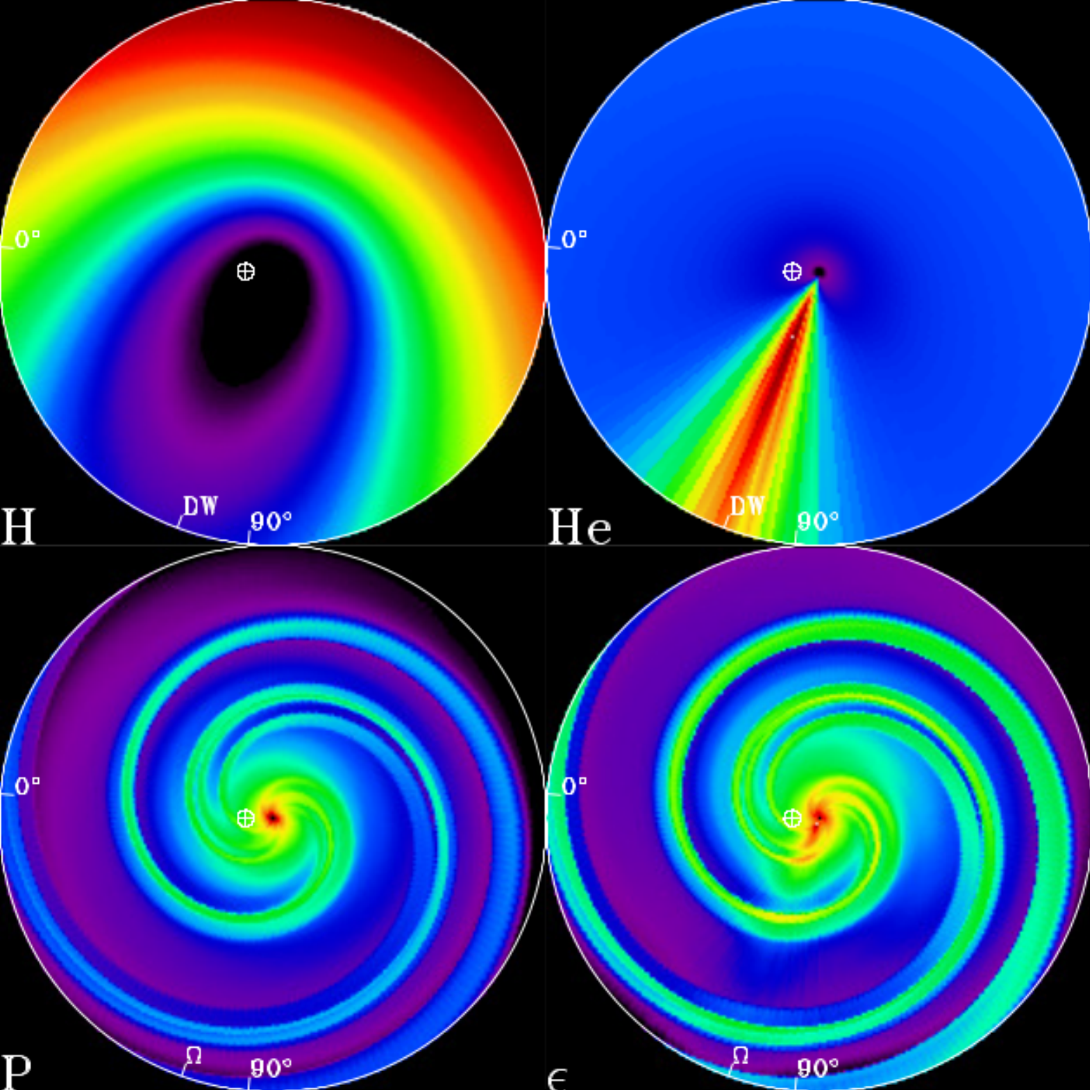}}
\caption{Critical quantities for the heliospheric model.
 {\bf Top Left: } The H distribution in the Earth's orbital plane in the inner solar system. The circle containing data is centered on the Sun and has a radius of 10 au. The downwind direction is marked as are the ecliptic longitudes of $0\arcdeg$ and $90\arcdeg$. The location of the Earth this simulation step is marked. The color scale is linear and runs from a minimum at black/purple (0 cm$^{-3}$) to a maximum at red/white ($6.07\times10^{-2}$ cm$^{-3}$).
 {\bf Top Right: } The He distribution, showing the He focussing cone. The projection and key is the same as in the previous panel. The color scale is linear and runs from 0 to $5.2\times10^{-2}$ cm$^{-3}$.
 {\bf Bottom Left: } The proton density in the solar wind. Here the longitude of the ascending node of the solar equator is marked. The color scale is logarithmic and runs from $10^{-2}$ cm$^{-3}$ to $10^{3}$ cm$^{-3}$.
 {\bf Bottom Right: } The relative soft X-ray emissivity, $\epsilon=(n_H+\mathcal{F}n_{He} )n_{swp}v_{rel}$ assuming that the H and He charge-exchange cross-sections have a ratio of 2:1 The color scale is logarithmic.
\label{fig:plane_demo}}
\end{figure*}

The {\it Local Heliospheric SWCX } is due to the interaction of the solar wind with the neutral interstellar medium (ISM) flowing through the solar system. Due to ionization and solar wind pressure, the H component of the ISM is mostly excluded from the inner solar system, leaving predominately the He component. The emissivity declines with distance from the Sun, due to the $R^{-2}$ dependence of the solar wind density, so the heliospheric emission will be dominated by more local emission. Although the solar wind is highly variable, it is, in the absence of a strong coronal mass ejection (CME), strongly ordered in the Parker spiral pattern. Thus, the local heliospheric emission is temporally variable and dependent on the look direction, but not so strongly as the magnetospheric emission. A simulation of the soft X-ray emissivity in the Earth's orbital plane is shown in Figure~\ref{fig:plane_demo}.

The {\it Distant Heliospheric SWCX } is due to the solar wind interacting with both interstellar H and He. The pathlengths are long, so the time variability is strongly reduced, but should reflect the solar cycle. Due to the roughly parabolic shape of the heliosphere, the strength must depend upon the look direction. The ``upwind'' direction is roughly $\ell=5\fdg1,b=19\fdg6$ \citep{lea2005} which suggests that the ISM flow is due both to the solar peculiar motion as well as either radial gas motion in the Galaxy or expansion from the Loop I superbubble \citep[see the review by][]{frisch2011}. In the following, this component will be taken as temporally constant.

For completeness, it should be noted that there will be SWCX due to the heliopause, but this will be a minor component. Using the \citet{dk2006} model for the distribution of interstellar neutrals in the heliosphere and the O$^{+7}$ cross-section therein, one can show that the fraction of O$^{+7}$ ions reaching a nominal heliopause at $\sim100$ au is $\sim50$\%. If one makes the (unlikely) assumption that all the remaining ions charge-exchange in the heliopause, the ratio of heliopause emission to the total heliospheric emission is $\sim0.03$. Assuming cross-sections of a similar order of magnitude for all of the species emitting in the \oqkev\ band, the heliopause is not a significant contributor to the heliospheric SWCX. 

Recent efforts have attempted to determine which of the temporally variable components dominates the SWCX ``contamination'' of \xmm\ or \chandra\ observations \citep{ks2008,csr2011,hs2012,wargelin2014}. The results have been equivocal and suggest that both components of the emission {\it can} be important in different situations. The simulations required to determine the SWCX production factor bring new insights to this issue and will be discussed below.  

\subsection{Methodological Overview}

The SWCX emission in a single emission line is given by the integral along the line of sight:
\begin{equation}
F = \int_{0}^{\infty} (n_{n} n_{swp} v_{rel} \langle \sigma \rangle f b) \frac{d\Omega}{4\pi} dl
\end{equation}
where $n_{n}$ is the density of neutral targets,
$n_{swp}$ is the solar wind proton density 
$v_{rel}$ is the relative velocity of the neutrals and the ions,
$f$ is the ratio of the density of the ion producing the line to the proton density,
$\langle\sigma\rangle$ is the velocity-weighted interaction cross-section,
and $b$ is the fraction of interactions that produce a photon in the line of interest.
It is necessary to use an averaged interaction cross-section since the interaction cross-section is velocity dependent. However, the velocity dependence is generally small over the range of velocities found in the solar wind, so the uncertainty introduced by the need to weight by the velocity distribution is generally on the order of the uncertainty in the cross-section itself \citep[see, for example, the catalogue of cross-sections in][]{bodewits}.
The relative velocity of the ions and the neutrals is given by
\begin{equation}
v_{rel}=(v_{swp}^2+v_{therm}^2)^{\frac{1}{2}}
\end{equation}
where $v_{therm}\sim3kT/m_p$ for the solar wind. 

Semi-empirical models of the distribution of the neutral material within the line of sight exist \citep{fahr1971,hodges1994}, as do semi-empirical models for the solar wind \citep{odstrcil2003} and measurements of the solar wind at the Earth. Thus, one should be able to calculate the SWCX emission within a given line of sight at a given time, though only to within the rather large uncertainty inherent in the models. However, the primary problem is that the interaction cross-sections, $\langle\sigma\rangle$, and branching ratios, $b$, are poorly measured, particularly in the energy range of astrophysical interest \citep[see the extensive discussion of alternative methods in][]{smith2014}. However, some important single lines, such as \ion{O}{7} and \ion{O}{8} have been somewhat characterized. The vast number of faint blended lines that contribute to the \oqkev\ band, however, have not. Since, with the notable exception of the Diffuse X-ray Spectrometer \citep{dxs} and X-ray Quantum Calorimeter \citep{xqc}, the \oqkev\ band has not been observed with sufficiently high spectral resolution to identify line complexes, let alone individual lines, we must deal with the aggregate of all of these lines. In such a case the \oqkev\ band emission can be written as
\begin{equation}
F = \int_{0}^{\infty} (n_{n} n_{swp} v_{rel} \varsigma) dl
\end{equation}
where
\begin{equation}
\varsigma =   \frac{d\Omega}{4\pi} \sum\limits_{i}^{} \langle \sigma_i \rangle f_i b_i 
\end{equation}
is the {\it production factor} for the energy band, summing over all of the different species emitting in the band.
It is the aim of this paper to determine $\varsigma$ for the \rosat\ \oqkev\ band. The above expression is often simplified to
\begin{equation}
F =  \varsigma Q
\end{equation}
where
\begin{equation}
Q \equiv  \int_{0}^{\infty} (n_{n} n_{swp} v_{rel}) dl.
\end{equation}
This expression assumes that the ionization structure is constant along the line of sight. Although such an assumption is surely incorrect, we will see the degree to which it is immaterial; the magnetosheath is smaller than the spatial/temporal scales upon which we can currently measure the changes in the ionization structure while the heliosphere is so large that the variations disappear when integrating along the line of sight.

Our primary source for \oqkev\ data is the \rosat\ All-Sky Survey (RASS). In the course of constructing the maps from the survey, it was realized that there was a temporally variable component of the soft X-ray background whose frequencies ranged from hours to days. Longer-term variations could not be characterized with the RASS, while shorter term variations were identified with other phenomena, such as aurorae, and discarded. \citet{mf_swcx1998} first suggested that these Long-Term Enhancements \citep[LTE,][]{sea1994} were correlated with the solar wind and \citet{crs2001} demonstrated the correlation more concretely. Since the foreground rate towards the dark Moon was roughly consistent with the LTE rate at the time of the observation \citep{collier2014}, it would seem that the LTE were due to SWCX in the magnetosheath where the neutral species charge-exchanging with the solar wind is predominately H. In this case, one should be able to use an MHD model of the magnetosphere for the spatial distribution of the solar wind ions in the near Earth environment, a model of the H distribution in the exosphere, and measurements of the local solar wind flux to calculate the quantity
\begin{equation}
\int_0^{\infty}(n_{en} n_{swp} v_{rel})dl \equiv Q_M.
\end{equation}
Here, $n_{en}$ is the density of exospheric neutral H and $Q_M$ is the $Q$ for the magnetosheath alone.
The ratio between the observed LTE rate and $\frac{d\Omega}{4\pi}Q_M$ should yield the production factor, $\varsigma$, for the \oqkev\ band for solar wind charge-exchange with neutral H.

There are, of course, a number of complications to be considered. The primary complication is the local heliospheric SWCX emission, which is due to the interaction of the solar wind with (predominately) neutral interstellar He flowing through the solar system. If the LTE flux were correlated with the local heliospheric SWCX emission, we would need to remove the variation due to the local heliosphere from the LTE light-curve before determining the production rate from the magnetosheath. In the following, we will use a MHD model of the solar wind and a semi-empirical model of the interplanetary distribution of neutral H and He to measure the correlation between the local heliospheric SWCX emission and the local solar wind flux. We will show that the local heliospheric emission provides a negligible contribution to the LTE rate.

To address the issue of the existence of the LHB requires more than the H production factor (derived here), and the He production factor derived by \citet{dxl2014}, it requires careful understanding of the shape of the heliosphere and the extent of emission from the heliopause. Both of these issues are beyond the scope of this work and will be addressed in the future.

In the following, \S2 describes the LTE data, the solar wind data, the MHD model of the magnetosphere, the MHD model of the solar wind for $R<10$ au, the semi-empirical model of the interplanetary distribution of neutral H and He, and the semi-empirical model of the Earth's exosphere. Section 3.1 describes the correlation between the local solar wind flux ($n_{swp}v_{swp}$) and the magnetosheath emissivity while \S3.2 describes the correlation between the \rosat\ LTE and the local solar wind flux. Section 3.3 demonstrates that the local heliospheric SWCX emission does not contribute significantly to the $n_{swp}v_{swp}$-LTE correlation. The discussion in \S4 explores the implications of the SWCX production factor. That section also discusses the implications of the heliospheric correlation analysis for observation and mission planning, as well as the question of the relative importance of the magnetosheath and heliosphere to SWCX ``contamination''.
  
\section{Data}
\subsection{X-ray Data}

\begin{figure*}
\center{\includegraphics[width=5.5cm,angle=90.0]{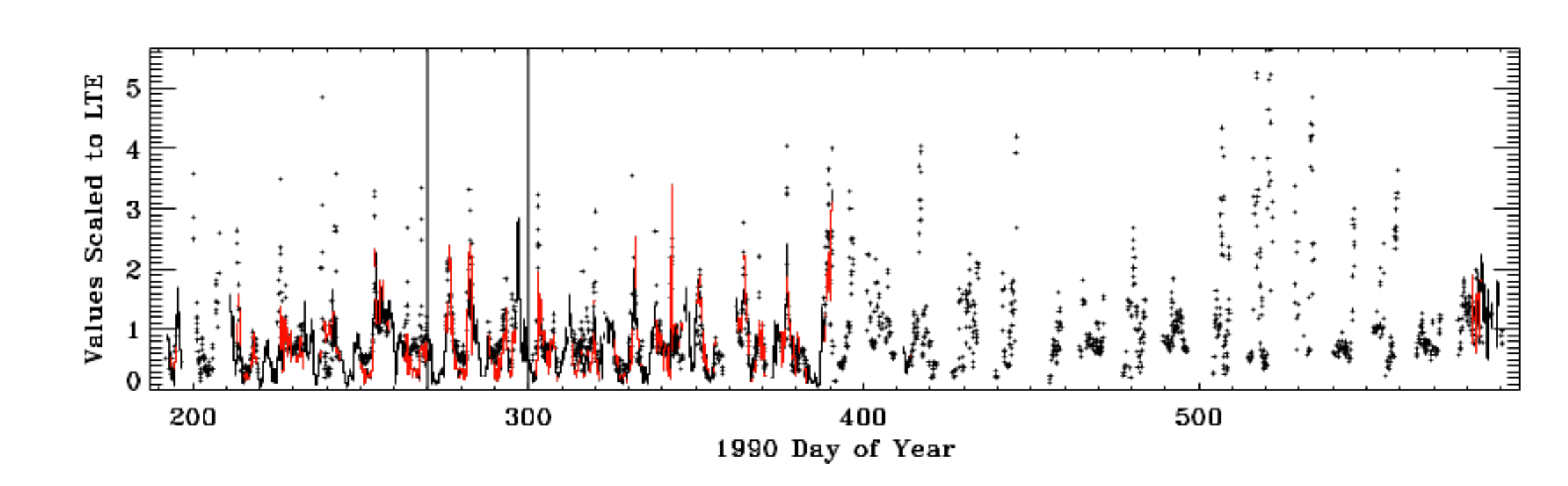}}
\caption{The \oqkev\ LTE flux ({\it solid lines}) and the solar wind flux ({\it crosses}) as a function of date. The LTE flux is in units of \rosat\ count s$^{-1}$ FOV$^{-1}$. The solar wind flux from \imp\ has been scaled to match the LTE flux. Periods of LTE data for which there is also solar wind data are plotted in {\it red}. The vertical lines denote the interval considered by \citet{crs2001}.}
\label{fig:c_lc}
\end{figure*}

{\it Orbit: } 
\rosat\  was launched 1 June 1990 into a nearly circular 96-minute orbit with an inclination of 53$\arcdeg$. The RASS observations began 11 July 1990 and continued, with some interruptions, for roughly six months. During the All-Sky Survey observations, the spacecraft was spun on an axis perpendicular to the optical axis. The spin axis was set to be within $\sim13\arcdeg$ of the Sun and precessed at $\sim4\arcmin$ per orbit, nearly the same rate as the Earth revolves about the Sun. The spin period was nearly the same as the spacecraft's orbital period about the Earth. Thus, over the course of a single orbit, the spacecraft observed a great circle roughly perpendicular to the Earth-Sun line passing through the ecliptic poles\footnote{Since the orbit precesses by $\sim18\arcmin$ per orbit while the plane of the look direction is fixed in GSE coordinates, the relation between location of the spacecraft and the look direction is continuously changing. The one constant is that the look direction is at  roughly +(-)GSE-Z when the spacecraft is at its maximum(minimum) GSE-Z. Similarly for GSE-Y. GSE-X however is more complicated. It should also be noted that the spin vector was reversed on a number of occasions to keep the spacecraft from scanning the Earth.}. Over the course of the bulk of the survey, the spin axis drifted to within $\sim7\arcdeg$ of the Sun. 

{\it LTE Isolation: } 
The \rosat\ Position Sensitive Proportional Counter (PSPC, the instrument with which the RASS was executed) FOV had a diameter of 2\arcdeg, while successive scans were offset by roughly 4\arcmin. Thus, each location on the sky could be observed for up to 30 consecutive orbits at the ecliptic plane and for many more orbits toward the ecliptic poles. Comparison of measurements of the same location on the sky from successive orbits allowed the detection and measurement of the time-varying background component called the Long-Term Enhancements (LTE). The details of the identification and measurement of the LTE in the RASS are given in \citet{sea1994,rass1995}. This time-varying component was removed from the RASS, although some residual LTE contamination remains and can be identified as stripes along lines of constant ecliptic longitude. Any non-temporally variable component due to the SWCX remains in the RASS, and it is the strength of this component that has been problematic and controversial \citep[compare, for example,][]{rl2004,ir2009,dk2009,wea1999}. 

{\it Binning: } 
The \oqkev\ band, C band, or R12 band LTE light-curve is shown in Figure~\ref{fig:c_lc}. The light-curve is discontinuous because the observations used for the RASS were not entirely continuous, mostly due to passage of the satellite through the radiation belts and the South Atlantic Anomaly. A portion of the \oqkev\ LTE light-curve, roughly days 270 to 300, was used by \citet{crs2001} to demonstrate that the LTE were correlated with the solar wind flux (as measured by \imp\ ), and thus were due to SWCX. For this work we will use the entire light-curve, though the contributions past day 400 do not contribute significantly. The data were originally binned into $\sim5$ minute intervals with a mean count rate of 0.7 count s$^{-1}$. No uncertainty was assigned in the original measure of the LTE rates and it is clear, given the way the values were derived, that the systematic uncertainties must be larger than the Poisson statistics. We have rebinned the data into a uniform 95-minute spacing and have taken the uncertainty of each bin to be $\sigma/\surd\overline{N}$ where $\sigma$ is the RMS variation of the data points contributing to an individual bin. Bins containing fewer than two data points were discarded. The bin spacing was chosen so that each bin contains roughly a single orbit of data and that adjacent data points cover nearly the same part of the sky. Thus, any systematic effects due to the cosmic background will vary by a small amount over a day and a half, rather than producing large variations between adjacent data points.

\begin{figure*}
\includegraphics[width=5.5cm,angle=90.0]{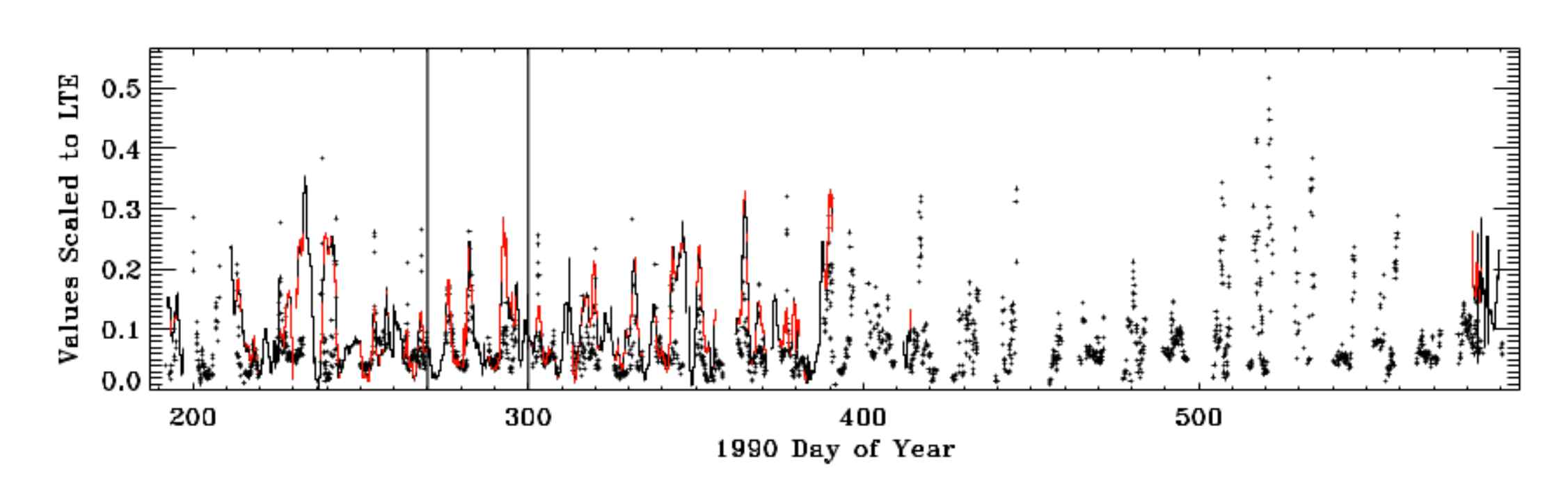}
\caption{The \tqkev\ LTE flux ({\it solid lines}) and the solar wind flux ({\it crosses}) as a function of date. The LTE flux is in units of \rosat\ count s$^{-1}$ FOV$^{-1}$. The solar wind flux from \imp\ has been scaled to match the LTE flux. Periods of LTE data for which there is also solar wind data are plotted in {\it red}. The vertical lines denote the interval considered by \citet{crs2001}.}
\label{fig:m_lc}
\end{figure*}

{\it Cusp Removal: } 
For the bulk of each orbit, the line of sight passed through the flanks of the magnetosheath where the SWCX emission is relatively low. However, in some seasons, the line of sight scanned near the magnetospheric cusps, where SWCX should be strongly elevated, as shown in Figure~\ref{fig:ms_demo}. The cusps are relatively narrow throats of magnetic field lines just poleward of the last closed field line on the Earth's day side where the Earth's magnetic field lines are open and thus they are the regions where the solar wind plasma can have direct access to the upper atmosphere. The cusps are threaded by the field lines undergoing or having recently undergone reconnection, and the particle population is determined by kinematic effects not included in MHD theory. Due to this limitation of MHD models, the density structures within the cusps are not properly characterized, and so need to be excluded from the analysis. Typically, the cusp extends 3-4 degrees in magnetic latitude \citep{zhou2000, palmroth2001} and 2-3 hours in longitude \citep{merka2002} at a given altitude. Due to the curved magnetic field lines threading the cusp, a low altitude spacecraft can observe lines of sight through the cusp over $\sim15\arcdeg$ in latitude, meaning that the spacecraft will scan across it in less than five minutes. In the RASS processing, short-term enhancements (STE) lasting just a few minutes when the line of sight passed near the Earth's pole were attributed, at the time, to auroral emission but may have been due to the cusps. These STE periods were removed from processing, and thus are not included in the LTE data. 

Removal of the STE periods does not guarantee the removal of cusp periods. We have attempted to identify periods when the line of sight passes through the cusp. This is not a trivial matter as the location of the cusp varies with diurnal motion, the annual motion, the direction of the interplanetary magnetic field, and the strength of the solar wind flux. We have used the \citet{ts04} model of the Earth's magnetic field to determine the magnetic field geometry along the line of sight, that is, whether the magnetic field lines through which the lines of sight pass are closed, open, or interplanetary. 

We first compared the LTE light-curve with a representation of the magnetic field geometry along the line of sight and saw no obvious correlation. We further checked time intervals for which viewing through the cusp was most likely: when the spacecraft was on the dayside and the line of sight was within $2\arcdeg$ of the GSE-Z axis, or when we might expect to be under(over) the northern(southern) cusp and to be looking up(down) through it. For each time interval meeting these criteria and for which there were both LTE and solar wind data, we calculated the difference, $\Delta_2$, between the LTE rate at that time and the LTE rate at $\pm10$ minutes. Since the scan rate is $~3\fdg75$ minute$^{-1}$, the cusp should fall in the center bin while the bins 10 minutes away should be free of the cusp. The histogram of the values of $\Delta_2$ is peaked at -0.05 count FOV$^{-1}$ s$^{-1}$, has a $\sigma=0.25$ count FOV$^{-1}$ s$^{-1}$. There is  a slight positive asymmetry, with an excess of events with 0.5 to 1.0 FOV$^{-1}$ s$^{-1}$, and  with a values extending to 1.3 count FOV$^{-1}$ s$^{-1}$. Time intervals with particularly high $\Delta_2$ did not show any particular distribution in spacecraft coordinates. Since STEs were removed from the LTE data and our search for enhancements at times likely to contain cusp emission did not reveal systematic increases in emission, we can confidently state that the residual contribution by the cusp to the LTE light-curve is small.

{\it The \tqkev\ Band: } The M band, or R45 band LTE light-curve is shown in Figure~\ref{fig:m_lc}. The \tqkev\ band data were treated in the same manner as the C-band data. Comparison to Figure~\ref{fig:c_lc} shows that the LTE in the two bands are rather different. The differences are due to the SWCX spectrum. Although the details of the SWCX spectrum are poorly understood, the overall shape of the spectrum is set by the ion abundances in the solar wind and the location of their principal recombination lines. The \tqkev\ band is dominated by the \ion{O}{7} and \ion{O}{8} lines with occasional significant contributions from \ion{Ne}{9} and \ion{Mg}{11} \citep[see, for example,][]{cs2008}. Rather than a few strong lines, the \oqkev\ band contains a large number of faint lines from many different species. Thus, the \tqkev\ band light-curve strongly depends upon the relative abundance of oxygen (compared to protons) and the relative ionization states of that oxygen. Conversely, since the \oqkev\ band contains such a multitude of lines, none of which dominates the band, the abundance and ionization variations average out. 

\subsection{Solar Wind Data}

Solar wind data for the period of the RASS are somewhat sparse for two reasons. First, the only satellite monitoring the solar wind at that time was \imp  . Since \imp\ was launched to study not only the solar wind, but the magnetospheric boundary and the magnetotail, it was exposed to the free-flowing solar wind for only a portion of its orbit. \imp\ did not have a means of recording data, and telemetry coverage was only 65\% to 80\% in this period. Thus, the solar wind parameters are measured for only a portion of the period for which we have LTE data. The solar wind data themselves were drawn from the NASA OMNI compilation\footnote{ftp://spdf.gsfc.nasa.gov/pub/data/omni/high\_res\_omni/} as that data set has been cleaned of inappropriate data periods and other data artifacts, as well as time shifted in a standardized manner. From these data we derive the ``local solar wind flux'', that is, the solar wind flux, the product of the proton density and the proton speed, in the neighborhood of the Earth.

\subsection{Magnetospheric Simulations}

\begin{deluxetable*}{lcrrrrr}
\tablecaption{BATSRUS Runs}
\tabletypesize{\scriptsize}
\tablecolumns{7}
\tablewidth{0pt}
\tablehead{
\colhead{Name} &
\colhead{Version} &
\colhead{Dipole} &
\colhead{Start} &
\colhead{Step} &
\colhead{Number} &
\colhead{Label} \\
\colhead{of Run\tablenotemark{a}} &
\colhead{ } &
\colhead{Tilt} &
\colhead{Date} &
\colhead{Size} &
\colhead{of Steps} &
\colhead{} \\
\colhead{} &
\colhead{} &
\colhead{(deg)} &
\colhead{} &
\colhead{(min)} &
\colhead{} &
\colhead{}}
\startdata
\cutinhead{Northern hemisphere winter solstice}
SWPC\_SWMF\_052811\_2 & v20110131 & -32.50 & 2006/12/14 & 1 & 2190 & $\Diamond$ \\
Marc\_Kornbleuth\_120513\_3 & v20110131 & -31.60 & 2002/12/19 & 10 & 37 & $\times$ \\
Marc\_Kornbleuth\_111213\_3 & v20110131 & -26.26 & 2004/11/06 & 10 & 6 & $\Box$ \\
Ankush\_Bhaskar\_050612\_1 & v20110131 & -25.81 & 2012/01/08 & 5 & 139 & + \\
\cutinhead{Equinox}
Yaireska\_Collado-Vega\_112812\_1 & v20110131 & -2.50 & 1991/03/24 & 1 & 601 & $\Diamond$ \\
SWADESH\_PATRA\_040412\_2 & v8.01 & -2.06 & 2002/03/24 & 4 & 510 & $\Box$ \\
Lur\_Zizare\_032112\_1 & v8.01 & -0.45 & 2012/03/12 & 4 & 511 & $\times$ \\
tao\_huang\_082814\_6 & v20140611 & -0.17 & 2008/02/23 & 4 & 76 & $\triangle$ \\
Ilja\_Honkonen\_073014\_2 & v20130129 & 2.15 & 2005/08/22 & 4 &1258 & + \\
\cutinhead{Northern hemisphere summer solstice}
Yuni\_Lee\_012610\_3 & v8.01 & 19.00 & 2001/06/01& 5 & 132 & $\triangle$ \\
SWPC\_SWMF\_060411\_6 & v20110131 & 29.50 & 2005/05/14 & ~12 & 388 & $\cdot$ \\
Steven\_Snowden\_011212\_1 & v8.01 & 28.66 & 2006/06/04  & 5 & 337 & $\Box$ \\
Yaireska\_Colladovega\_091112\_1 & v20110131 & 28.88 & 2007/05/20 & 1 & 50 & $\Diamond$ \\
Chigomezyo\_Ngwira\_022014\_1 & v20130129 & 29.83 & 2013/06/28 & 1 & 1200 & $\times$ \\
Brian\_Walsh\_030413\_1 & v20130129 & 30.04 & 2000/07/15 & 1 & 338 & + \\ 
\enddata
\tablenotetext{a}{The name of the run as it appears on the CCMC run-on-request website.}
\label{tab:runs}
\end{deluxetable*}

We represent the ion density of the  magnetosheath with simulations using the  \bats\ (Block-Adaptive-Tree Solar Wind Roe-Type Upwind Scheme)\footnote{Simulation results have been provided by the Community Coordinated Modeling Center at Goddard Space Flight Center. The BATS$-$R$-$US model was developed by Dr. Tamas Gombosi at the Center for Space Environment Modeling, University of Michigan.} code \citep{toth2005,powell1999}.  \bats\ solves the three-dimensional MHD equations for the region surrounding the Earth, simulating the response of the magnetosphere to the temporally variable solar wind.  \bats\ is one of several standard MHD models of the magnetosphere in common use. Validation of these models by comparison with {\it in situ} spacecraft measurements is an ongoing effort among many institutions \citep[e.g.][among others]{raeder2003} and a new intercomparison of models is in progress (Collado-Vega \& Sibeck, in preparation, 2015).

It should be noted that the magnetosheath is sensitive not only to the solar wind flux and the IMF but also to the recent history of those parameters. This dependence is due, in part, to the time required for a solar wind impulse to move from the nose of the bowshock past the Earth ($\sim3$ minutes) and down the tail. The dependence is also due to the finite time for the magnetosphere to respond to changes in the solar wind. Thus MHD models are superior to analytic models, such as that of \citet{ssa1966}, which have been used for previous SWCX studies \citep{csr2011,ks2008}. In comparison to \citet{ssa1966} we find that \bats\ produces consistently different positions for the magnetopause and the bowshock (see \S4.2). \bats\ also models the asymmetry of the magnetosheath. 

Each \bats\ simulation is sampled at a predetermined cadence, usually every few minutes. Each sample produces proton density, speed, and temperature values for each point in a three-dimensional grid whose spacing is optimized to provide high resolution of strong gradients. One disadvantage of the \bats\ code is that it does not distinguish between protons originating in the solar wind, and those much lower energy protons originating in the plasmasphere closer to the Earth. However, the amount of solar wind plasma within the magnetosphere is relatively low \citep{christon1994}. Thus, one can isolate the solar wind protons by removing regions that lie within the Earth's closed field lines. This method has difficulties around the cusps, but as the cusps are problematic with MHD models, and have been removed from the LTE light-curves, this issue is not significant for this analysis. (However, see \S3.1.)

Running \bats\ for the roughly 200 days of the RASS would be a significant effort, particularly without continuous solar wind data. At the LTE sample rate of five minutes, we would accumulate 57600 samples, at roughly 75Mb each. Even sampling once per orbit would produce 3200 samples. While such sampling would be ideal, it is beyond the scope of this work. We have, however, accumulated a series of \bats\ simulations from other projects that cover a variety of solar wind conditions. The runs and their relevant parameters are listed in Table~\ref{tab:runs}. The validity of using representative simulations rather than dedicated simulations of the RASS epochs will become apparent in the analysis.

\subsection{The Model of the Exospheric Neutral Distribution}

The neutral density in the magnetosphere is taken from the model of \citet{hodges1994}. This static model was originally calculated for a variety of values of insolation as measured by f10.7\footnote{The 10.7 cm (2800 MHz) solar flux as measured by the Penticton observatory.} at both solstice and equinox. For any particular set of conditions and dates, we interpolate among the available model states. The model is valid from 1.05 \re to 9.75 \re ; at larger distances the model is extrapolated as $R^{-3}$. Comparison of the Hodges model calculations with Lyman $\alpha$ column brightnesses for the antisolar point \citep{oea2003} suggests that the Hodges model is compatible with the limited observations.

\subsection{Heliospheric Simulations}

We represent the ion density in the heliosphere with output from the ENLIL\footnote{Simulation results have been provided by the Community Coordinated Modeling Center at Goddard Space Flight Center. ENLIL was developed by Dusan Odstrcil at the University of Colorado at Boulder. ENLIL is not an acronym, despite its nonstandard capitalization, but the name of the Sumerian god of winds and storms.} code \citep{odstrcil2003}. ENLIL solves the three-dimensional MHD equations for the inner heliosphere. The inner boundary conditions of the solar wind density, speed, and temperature are derived from solar magnetograms. The inner boundary is taken to be 21.5 R$_{\sun}$ where the solar wind becomes supersonic.

The ENLIL simulation used for this work was specially requested of the CCMC. The first sampled time step of the simulation is 1990-09-28 12:18 UT. The simulation was sampled with a time step of 90 minutes (roughly the \rosat\ orbital period) for a total of 529 samples, or slightly over a month of simulation, corresponding to the period studied by \citet{crs2001}. The simulations were run over a standard region; 21.5 R$_{\sun} < R < 10$ au and $60\arcdeg < \theta < 120\arcdeg$ where $\theta$ is the angle from the north solar rotational pole. The restriction to within 30\arcdeg\ of the solar equator will not be of great concern for the correlation analysis to be done in \S\ref{sec:enlil}.

\subsection{The Model of the Heliospheric Neutral Distribution}
 
The neutral density in the heliosphere is represented by the model developed by \citet{dk2006} based on the classical hot model calculations of \citet{lbd1985}, \citet{lbk1985}, and \citet{lrv2004}. The model was calculated for solar maximum conditions such as those that would have been experienced during the \rosat\ All-Sky Survey observations. The ratio of photon pressure to gravitational force on H, $\mu$, is set to 1.46, while the total equatorial ionization rate (including CX with protons, photoionization and maybe electron impact) derived from the SWAN data \citep{quemerais2006} is $6.54\times10^{-7}$ s$^{-1}$. The H ISM parameters used for constructing these models are taken from \citet{lea2005}: upwind $(\lambda,\beta)=(252\fdg3,8\fdg5)$, $T=13,000$K, $v_0=21.0$ km s$^{-1}$, and $n_0=0.1$ cm$^{-3}$. The equatorial electron impact factor for He is given by the rates in \citet{rucinski1989} multiplied by 2.05, the factor estimated by \citet{lrv2004} to account for the increase during solar maximum. The He ISM parameters are taken from \citet{mobius2004} and \citet{witte2004}: upwind $(\lambda,\beta)=(254\fdg7,5\fdg2)$, $T=6300$K, $v_0=26.3$ km s$^{-1}$, and $n_0=0.015$ cm$^{-3}$. The H and He contributions are calculated separately as the ion-neutral interaction cross-sections are different for the two species.

\section{Analysis}

We will first determine the relation between the local solar wind flux, $n_{swp}v_{swp}$ and the $Q_M\equiv\int n_{en}n_{swp}v_{rel} dl$ for the magnetosheath. It is expected that this relation should be non-linear; as the solar wind flux increases, the bowshock is pushed closer to the Earth, where the neutral density is higher. We must determine what part of this relation is comparable to the relation between $n_{swp}v_{swp}$ and the \oqkev\ LTE rate. By combining the two relations, we then get a direct relation between $Q_{M}$ and the \oqkev\ LTE rate, and thus the \rosat\ \oqkev\ band flux.

\subsection{The Q$_M$ - Solar Wind Flux Relation}

\begin{figure*}
\center{\includegraphics[width=8.5cm,angle=0.0]{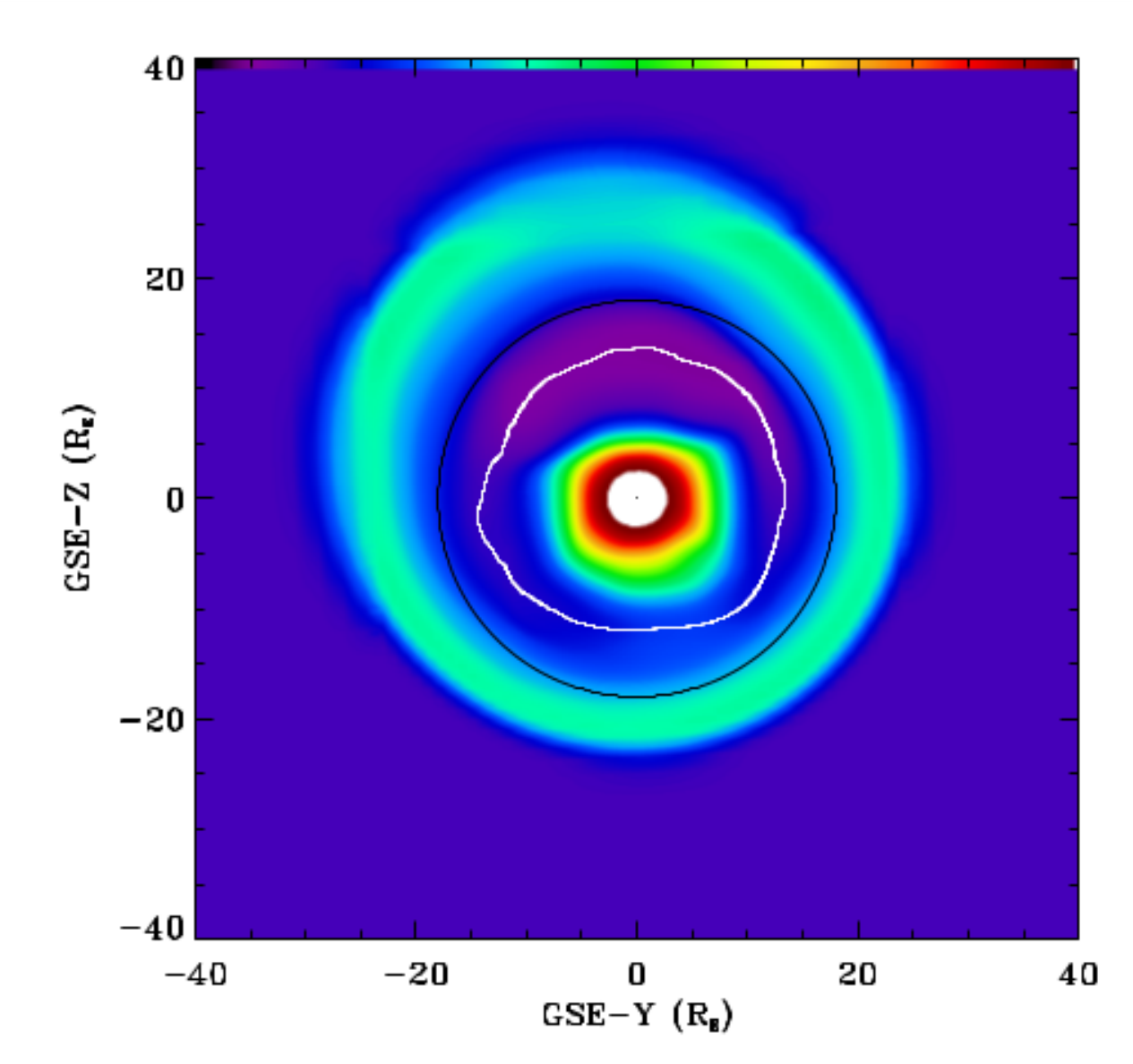}\includegraphics[width=8.5cm,angle=0.0]{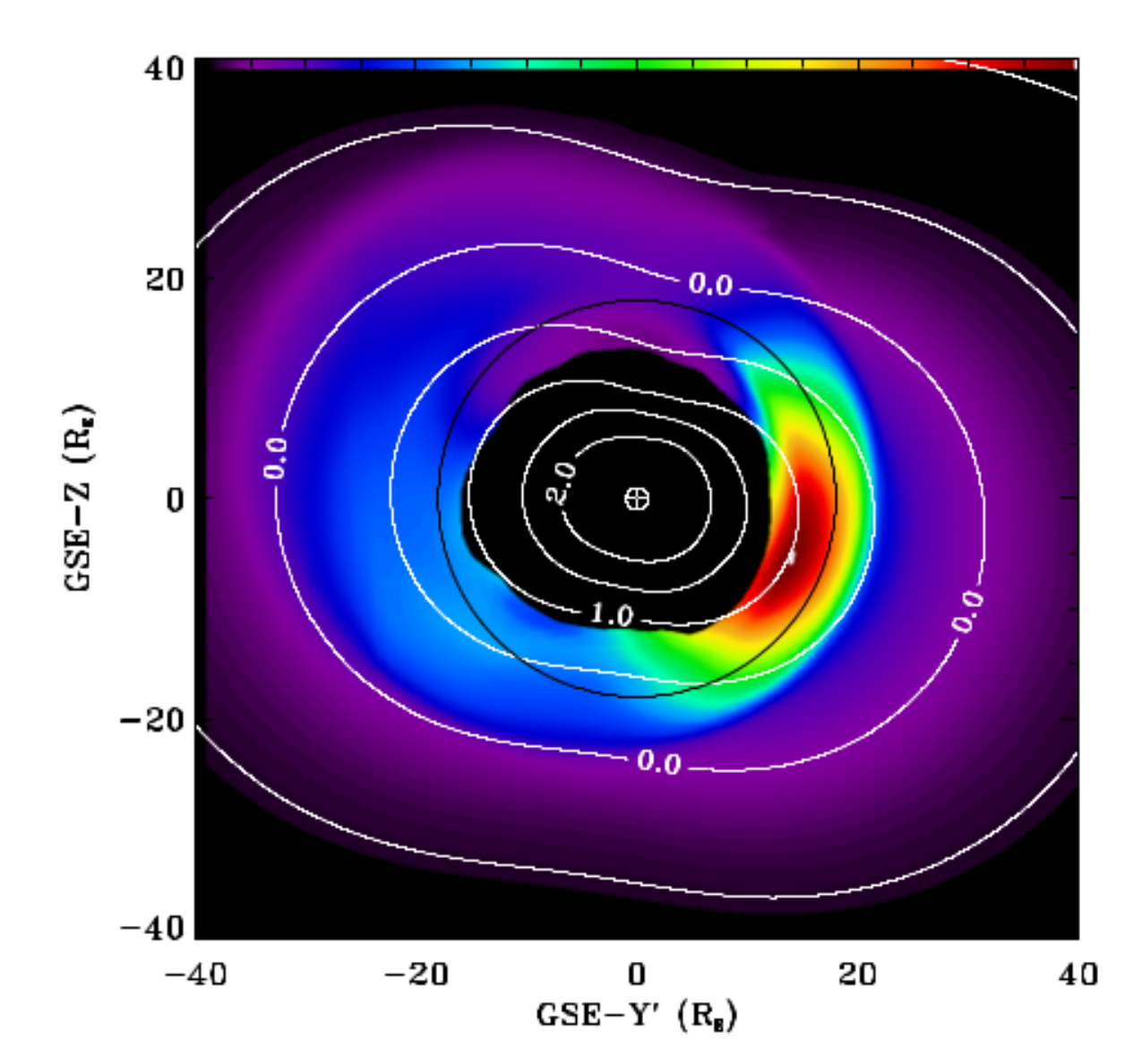}}
\caption{{\bf Left: } A cut through the magnetosphere in the GSE YZ plane. The colorscale traces the proton density and shows a peak around the Earth (the plasmasphere) and a peak at 20-25 R$_E$ (the magnetosheath). The {\it white circle} is centered on the Earth and has a radius of 18 R$_E$. The {\it black line} shows the boundary of the region excluded by the plasmasphere removal algorithm. The north ecliptic pole is up. {\bf Right: } A cut through the magnetosphere in the reference scan plane (Y$^{\prime}$Z) which is rotated 12\arcdeg about the GSE-Z axis from the GSE-YZ plane. The colorscale traces the SWCX emissivity. The maximum emissivity is $2.3\times10^9$ cm$^{-5}$ s$^{-1}$. The {\it black circle} is centered on the Earth and has a radius of 18 R$_E$. The contours trace the logarithm of the neutral density. The black region at the center is the plasmasphere as defined by the plasmasphere removal algorithm.
\label{fig:yz_demo}}
\end{figure*}

In order to understand the magnetosheath SWCX flux that \rosat\ ought to have observed during the All-Sky Survey, we sampled a number of magnetospheric simulations with lines of sight typical of the RASS. It is important to bear in mind that the LTE data with which we are to compare the results of simulations are binned to intervals of roughly one orbit. Thus we wish to extract from the simulations the equivalent to what \rosat\ would have observed over a single great circle scan. We begin with a ``reference'' scan, a great circle, centered on the Earth, passing through the ecliptic poles, and rotated $12\arcdeg$ about the GSE-Z axis from the GSE-YZ plane. For each time step of each simulation we calculated integrated emissivity, $Q_M$, for 360 lines of sight uniformly distributed over this great circle. From those we then calculated the mean integrated emissivity $\overline{Q_M}$ for the scan. Since this reference scan passes through the ecliptic poles (the GSE-Z axis) it avoids the cusps except for times around the solstices. Once we understand the behavior of $\overline{Q_M}$ for the reference scan, we will explore the effects on $\overline{Q_M}$ of using more realistic scans.

As noted in \S2.3, the \bats\ simulations do not distinguish between solar wind protons and the cold plasmaspheric protons, while the SWCX emission can be associated with only solar wind protons. We thus need to remove the plasmasphere from the simulation before calculating $Q_M$. Since these scans avoid the cusps, the proton density for any direction on a scan is peaked at the Earth, declines with increasing radius to some relative minimum, and then shows another peak at the magnetosheath. The plasmasphere can be crudely but rather efficiently defined as the region inside the relative minimum in proton density, though the plasmasphere by no means fills that region. Figure~\ref{fig:yz_demo} shows the proton density for a GSE YZ slice through one step of a simulation, as well as the boundary of the ``plasmaspheric'' region to be removed. This cleaning algorithm works well when the cusp does not enter the scan.

\begin{figure}
\center{\includegraphics[width=7.5cm,angle=0.0]{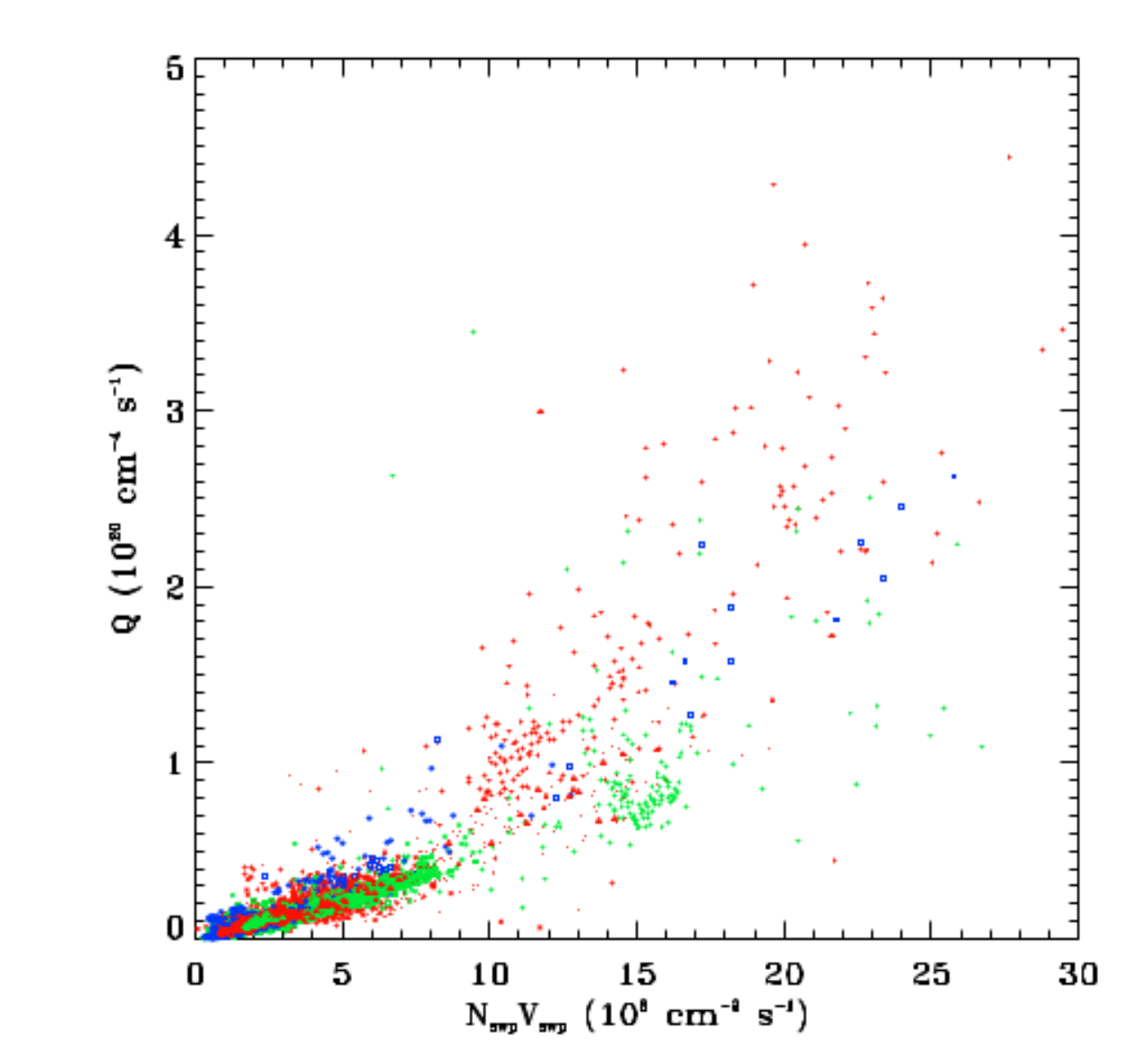}}
\vspace*{-0.75cm}
\center{\includegraphics[width=7.5cm,angle=0.0]{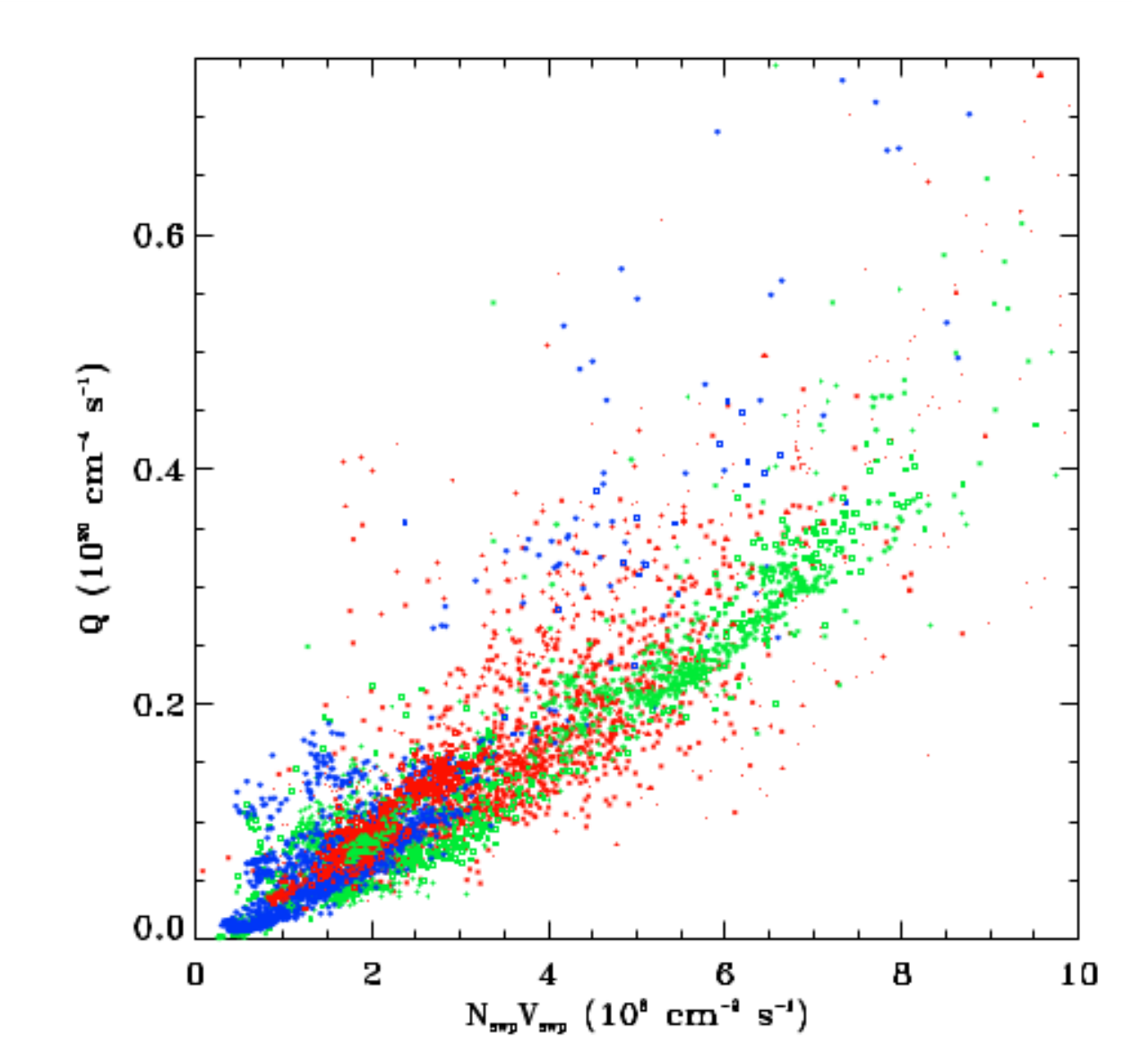}}
\vspace*{-0.75cm}
\center{\includegraphics[width=7.5cm,angle=0.0]{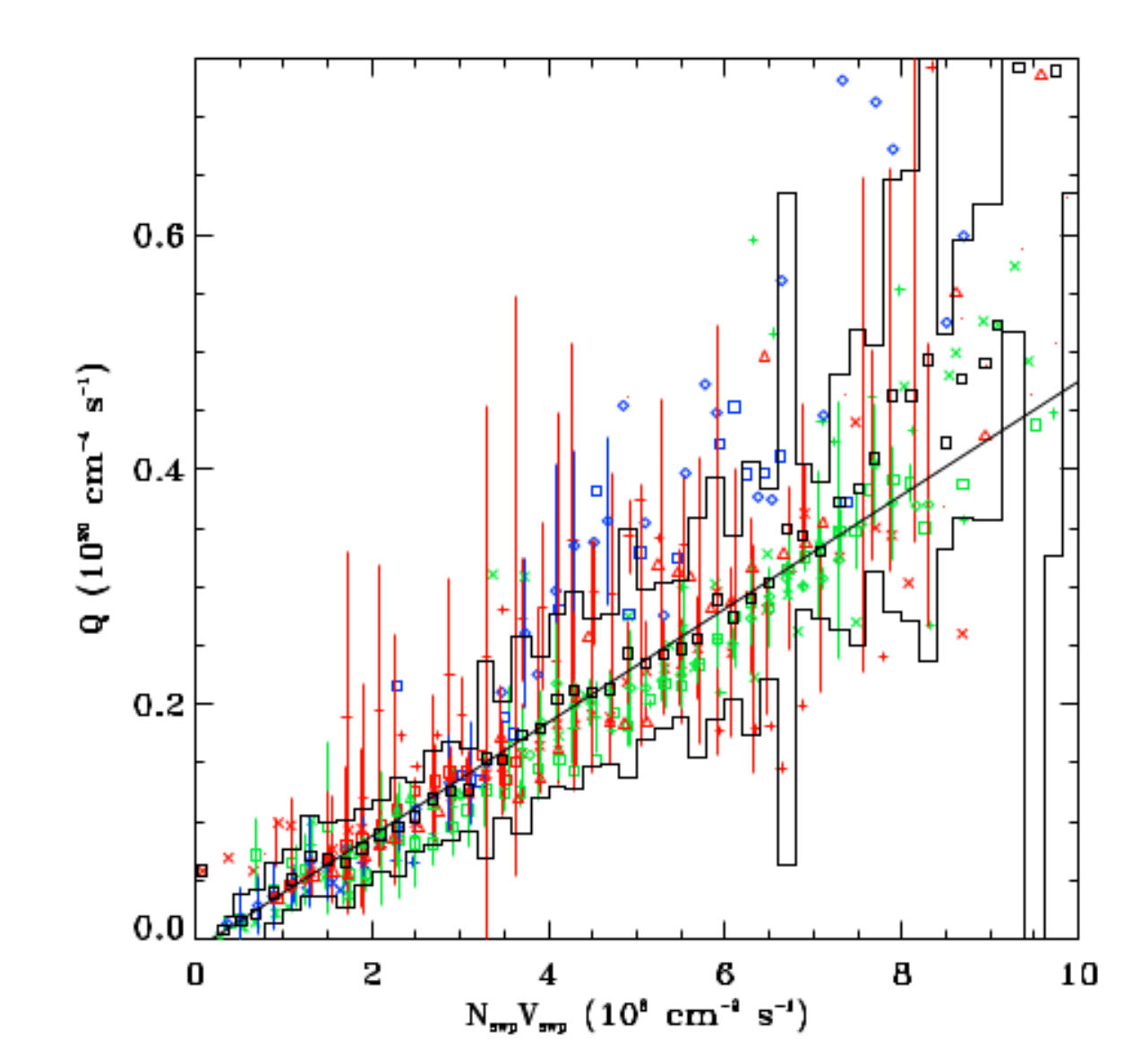}}
\caption{The $n_{swp}v_{swp}$-$\overline{Q_M}$ relation for the simulations in Table~\ref{tab:runs}. They are color-coded by season: {\it blue: } Northern hemisphere winter solstice, {\it green: } equinox, and {\it red: } summer solstice. Table ~\ref{tab:runs} provides a key to symbols where used. {\bf Top: } The bulk of the parameter space covered by the simulations. Each point is a single time step of one simulation. {\bf Middle: } A detail of the previous panel showing the parameter space covered by the LTE data. {\bf Bottom: } The data binned by $n_{swp}v_{swp}$. The {\it colored} data points are the binned data from individual simulations. The standard deviation of the data points in each bin is shown if there were more than five data points in the bin. The {\it black} points are the binned data from all the simulations. The black {\it histogram} shows the standard deviation for the black points. 
\label{fig:q_nv}}
\end{figure}

In the top two panels of Figure~\ref{fig:q_nv}, each plot point represents the $\overline{Q_M}$ for an individual time step of a simulation listed in Table~\ref{tab:runs}. The points have been color-coded by season and, where possible, symbol-coded by simulation. The top panel shows the bulk of the parameter space covered by the simulations, while the middle panel covers only the values of $n_{swp}v_{swp}$ found in the LTE data periods. The bottom panel shows the values binned into $\Delta n_{swp}v_{swp}=0.2\times10^8$ cm$^{-2}$ s$^{-1}$ wide bins. The colored symbols are points from individual simulations while the black points are summed over all the simulations. The standard deviations are shown for bins containing more that five values.

As one might expect, as the solar wind flux increases, the value of $Q_M$ increases non-linearly. The relation between $\overline{Q}$ and $n_{swp}v_{swp}$ becomes steeper as $n_{swp}v_{swp}$ increases. Below $n_{swp}v_{swp}=6\times10^8$ cm$^{-2}$ s$^{-1}$ the relation is relatively linear, or at least any non-linearity is obscured by the intrinsic variation. The bottom panel of Figure~\ref{fig:q_nv} shows a linear fit to the values with $n_{swp}v_{swp}<6\times10^8$ cm$^{-2}$ s$^{-1}$ and that fit seems adequate. The values do not deviate significantly from the linear fit until $n_{swp}v_{swp}\sim10\times10^8$ cm$^{-2}$ s$^{-1}$. There is a significant dispersion in $\overline{Q_M}$ at any given value of $n_{swp}v_{swp}$ for a single simulation, and clear systematic differences between simulations. At $n_{swp}v_{swp}>10\times10^8$ cm$^{-2}$ s$^{-1}$ there are fewer simulation points and the behavior of the relation becomes quite uncertain.

The first source of variation in $\overline{Q_M}$ at a given $n_{swp}v_{swp}$ is the inherent difficulty of defining the solar wind pressure acting upon the magnetosheath. The solar wind takes roughly three minutes to move from the nose of the bowshock to a plane passing through the center of the Earth. Each portion of the magnetosphere responds to the ambient pressure at its location, and the response propagates rapidly throughout the magnetosheath. Thus, setting a single characteristic $n_{swp}v_{swp}$ for a time step is inherently inaccurate. There is, as well, the difficulty of propagating strongly variable solar wind conditions from their point of measurement to the (moving) magnetopause or bowshock. Thus, there is an intrinsic error in setting $n_{swp}v_{swp}$, particularly when the solar wind conditions are changing rapidly. Since high $n_{swp}v_{swp}$ periods tend to be periods with strong, rapid variation in $n_{swp}v_{swp}$, the variation in $\overline{Q_M}$ increases with $n_{swp}v_{swp}$.  The simulations with slower changes in the solar wind show significantly less scatter. 

A second source of variation is the asymmetry of the magnetosheath, which depends not so much on the $n_{swp}v_{swp}$ of the solar wind, but upon the variation in the direction of the solar wind momentum vector and the direction of the interplanetary magnetic field. The asymmetry can be seen in Figure~\ref{fig:yz_demo}. Nor is the neutral density spherically symmetric; rather it is squashed at the poles, which leads to a rather asymmetric distribution of emission. Since the scan path is rotated by 12\arcdeg\ around the GSE-Z axis with respect to the YZ plane, it samples the magnetosheath at much closer distances on one side than it does on the other, which leads to even more extreme asymmetries. The asymmetry in the emission for a typical scan, after removal of the plasmasphere, is shown in the right-hand panel of Figure~\ref{fig:yz_demo}. The average of the integrated magnetospheric emissivity over a great circle scan, $\overline{Q_M}$, depends on the direction and amplitude of the rotation of the scan plane with respect to the underlying asymmetry. Thus, for any given value of  $n_{swp}v_{swp}$, one would expect a variation in $\overline{Q_M}$ due to the variation in the asymmetry. Using a representative set of simulations, reversing the sense of the rotation of the scan path with respect to the GSE-YZ plane changed the $\overline{Q_M}$ for individual time steps from -20\% to +10\%, with a mean change of -7\%. This variation might be reduced if the \citep{hodges1994} model overestimates the equatorial bulge of the exosphere.

A third source of variation is seasonal variation. As the Earth nods, the cusp latitude changes, bringing the cusps closer or further from the scan path. However, as indicated in Figure~\ref{fig:q_nv}, this variation seems to be small compared to the intrinsic differences between simulations.

The combined effect of all of these variations produce that scatter seen in Figure~\ref{fig:q_nv}. We have calculated the fractional variance as a function of $n_{swp}v_{swp}$. Using values for all of the simulations, we calculated the mean $\overline{Q}$  for $\Delta n_{swp}v_{swp}=0.2\times10^8$ cm$^{-2}$ s$^{-1}$ wide bins (the black boxes in the bottom panel of Figure~\ref{fig:q_nv}). The standard deviation of the values in each bin is shown as the black histogram. For $n_{swp}v_{swp}<8\times10^8$ cm$^{-2}$, the standard deviation in $\overline{Q}$, $\sigma_{\overline{Q}}$, is roughly 40\% of the mean ${\overline{Q}}$ and is relatively independent of $n_{swp}v_{swp}$.

By fitting the data in the last panel of Figure~\ref{fig:q_nv} we find that
\begin{equation}
\left[\frac{n_{swp}v_{swp}}{10^8}\right] = (0.037 \pm 0.189) + (20.68 \pm 0.41) \left[\frac{\overline{Q_M}}{10^{20}}\right]
\end{equation}
where $n_{swp}v_{swp}$ is in cm$^{-2}$ s$^{-1}$ and $\overline{Q_M}$ is in cm$^{-4}$ s$^{-1}$. The correlation is fitted from the values binned over all simulations (the black boxes in the bottom panel of Figure~\ref{fig:q_nv}) since the individual values derived from each time-step of each simulation have no intrinsic uncertainties. The measured standard deviation of each point was used as the uncertainties in the fit. We fit only the values below $n_{swp}v_{swp}<6\times10^8$ cm$^{-2}$ s$^{-1}$ where the relation is most clearly linear. We have used a fitting method \citep{ab1996} that minimizes the uncertainty perpendicular to the fit, and determines the uncertainties in the fit parameters using a bootstrap technique. Note that the uncertainty in the intercept is significantly larger than the intercept itself, thus the intercept is not significantly different from zero. 

Any intrinsic scatter in the $n_{swp}v_{swp}$-$\overline{Q_M}$ relation translates directly into a scatter in the $n_{swp}v_{swp}$-LTE relation. Besides the scatter explored above, there is also scatter due to the difference between the geometry of the reference scan and real \rosat\ scans. Over the course of an orbit, the spacecraft will move from $\sim1.09$ R$_E$ on the day side of the Earth to $\sim1.09$ R$_E$ on the night side of the Earth; the scan path moves correspondingly with respect to the reference scan. As the emission increases towards the sub-solar point of the magnetosheath, the more sunward the scan, the higher the emission. A real \rosat\ scan will sample the magnetosheath on both sides of the reference scan, but since the emission is not linear with distance from the sub-solar point, the effects do not cancel.

\begin{figure}
\center{\includegraphics[width=7.5cm,angle=0.0]{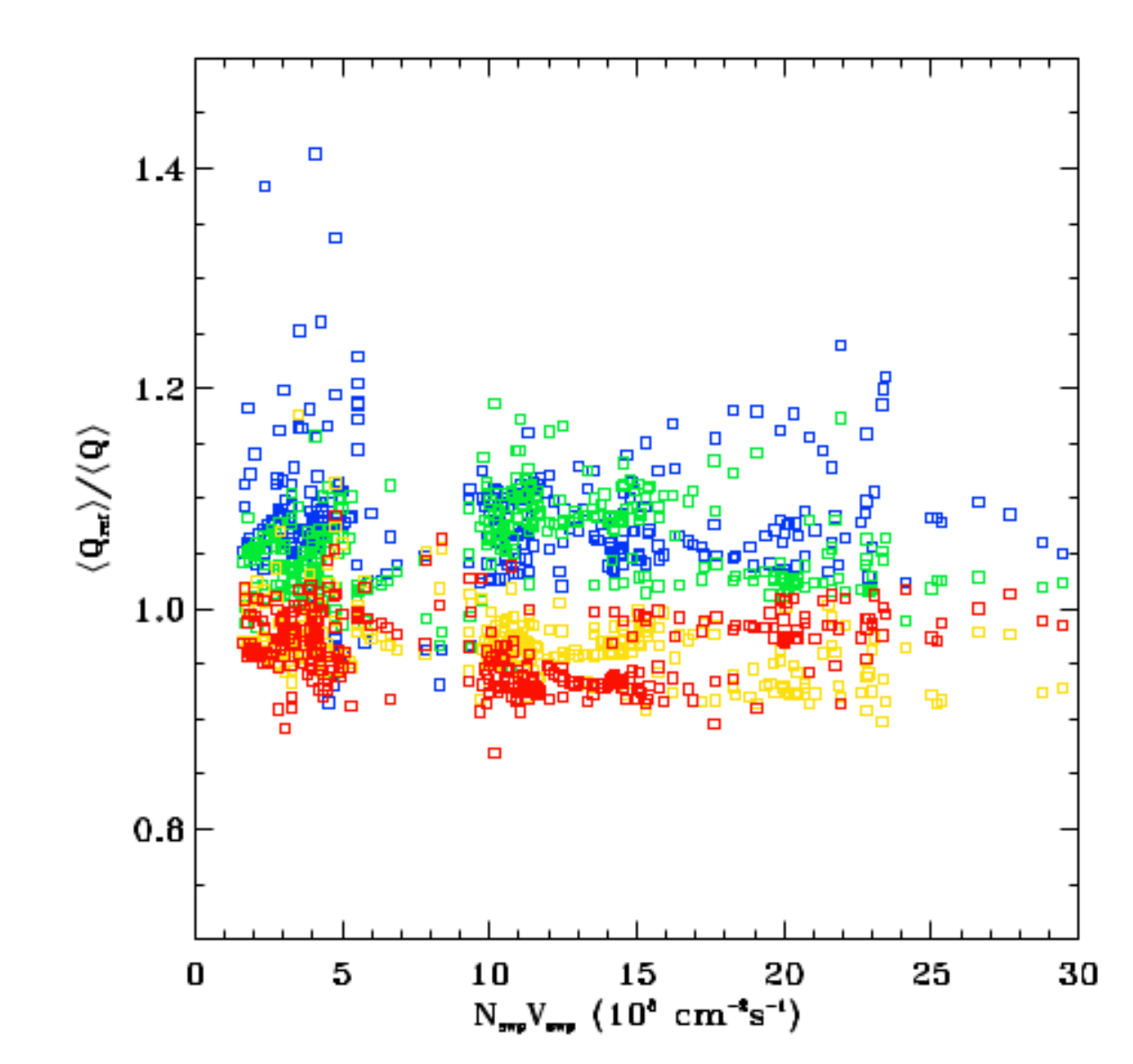}}
\caption{The ratio of the $\overline{Q_M}$ for the reference scan to the $\overline{Q_M}$ from four test scans, as a function solar wind flux. The test scans are color coded by the orbit location at which the scan is in the +GSE-Z direction: {\it blue: } noon, {\it green: } dusk, {\it yellow: } midnight, and {\it red: } dawn. The behavior is expected given that the day-side emission is stronger than night-side emission.
\label{fig:q_orbit}}
\end{figure}

To measure the difference between realistic scans and the reference scan, we constructed four test scans for a single (rather unrealistic) test orbit lying in the GSE-XY plane. Each test scan was constructed such that the line of sight is towards the north ecliptic pole when the spacecraft is at one of the cardinal directions, dawn, dusk, noon, or midnight. These four test scans sample, however sparsely, the possible range of orbit attitudes. We have compared the $\overline{Q_M}$ for these test scans to the $\overline{Q_M}$ for the reference scan. The result for a simulation near summer solstice when the solar wind flux spanned the bulk of the expected range is shown in Figure~\ref{fig:q_orbit}; the bulk of the differences between the real scans and the reference scan lie between -10\% and +20\%.

\subsection{The LTE - Solar Wind Flux Relation}

\begin{figure*}
\center{\includegraphics[width=8.0cm,angle=0.0]{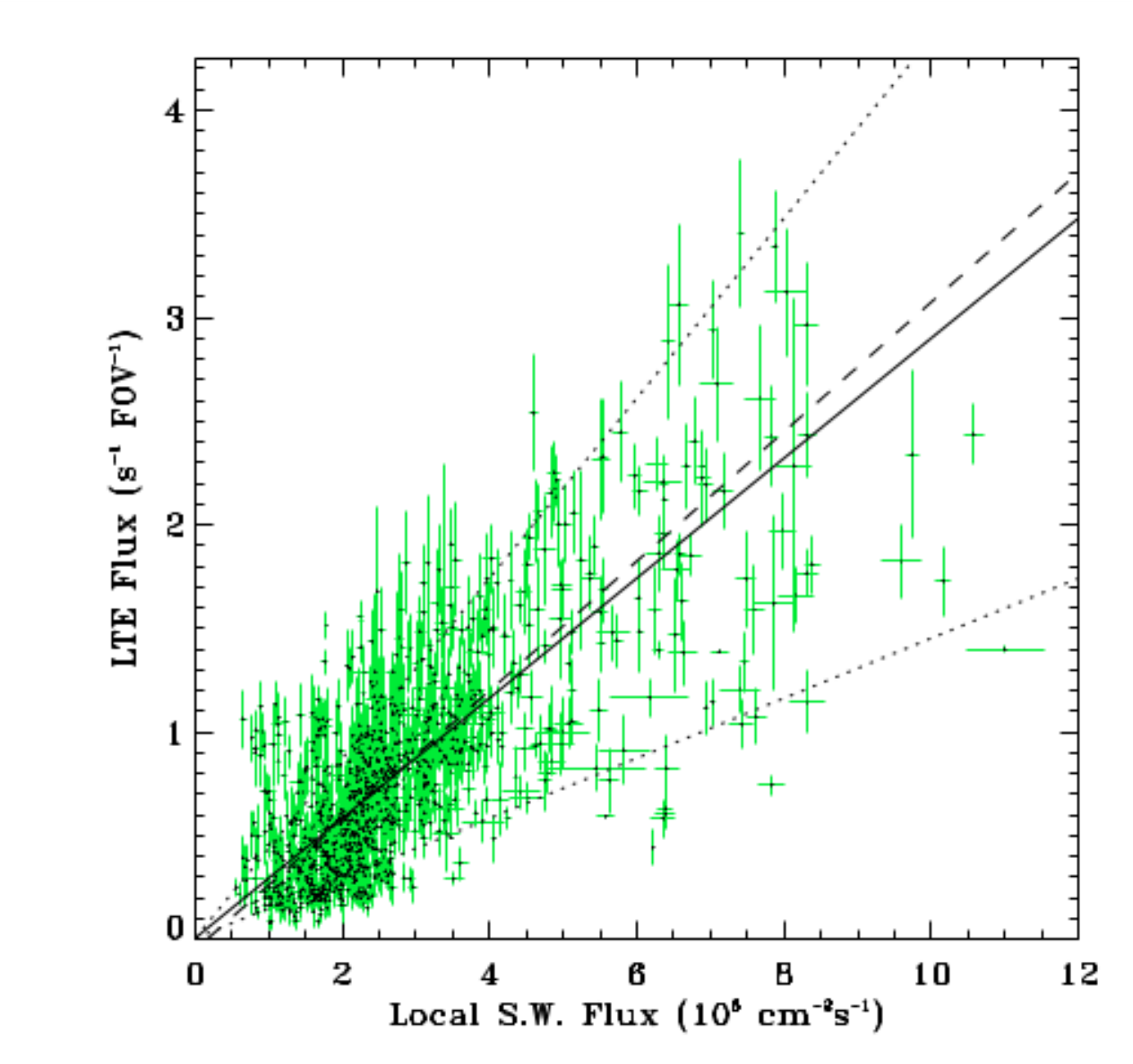}\includegraphics[width=8.0cm,angle=0.0]{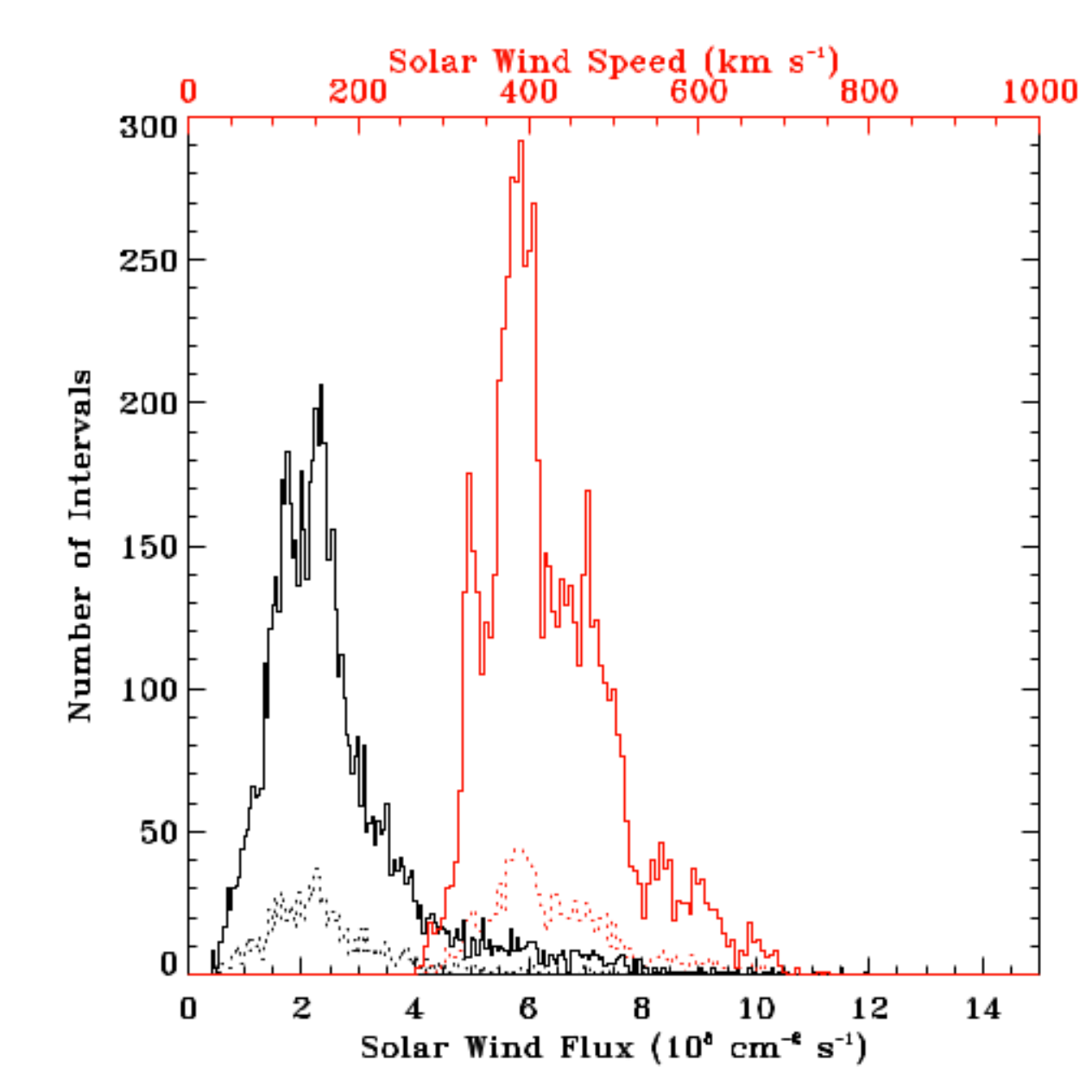}}
\caption{{\bf Left: } The \oqkev\ LTE flux as a function of the local solar wind flux. The {\it solid line} is the best fit of (LTE$\mid n_{swp}v_{swp}$), the {\it dashed line} is the best fit minimizing the distance orthogonal to the fitted line, and the {\it dotted lines} show the relations with 0.5 and 1.5 times the best fit slope. {\bf Right: } Histogram of the solar wind flux ({\it black}) and solar wind speed ({\it red}) values for the time periods with LTE data. The {\it dotted lines} are binned by orbit while the {\it solid lines} are for individual five minute intervals.
\label{fig:c_nv}}
\end{figure*}

The correlation of the \oqkev\ band LTE flux with the local solar wind flux is shown in Figure~\ref{fig:c_nv} (Left)  where, as in the light-curve, the data have been binned in 95 minute bins. The uncertainties shown are the RMS variations of the data being binned. As can be seen from the histogram of the solar wind data for the periods for which there is LTE data (Figure~\ref{fig:c_nv}) the bulk of the data is for $n_{swp}v_{swp}<5\times10^8$ cm$^{-2}$ s$^{-1}$. Above that value of the local solar wind flux the data become increasingly sparse, tailing off by $~10^9$ cm$^{-2}$ s$^{-1}$. The bulk of the LTE observations were made when $n_{swp}v_{swp}$ was well within the range for which the $n_{swp}v_{swp}$-$\overline{Q_M}$ relation is linear, and only the most extreme values fall in the regime where the linearity is dubious. We have fitted the correlation for $n_{swp}v_{swp}<6\times10^8$ cm$^{-2}$ s$^{-1}$using the bootstrap method of \citet{ab1996}. We find
\begin{equation}
LTE = (0.0012 \pm 0.0353) + (0.291 \pm 0.014) \left[\frac{n_{swp}v_{swp}}{10^8}\right]
\end{equation}
where LTE is in count s$^{-1}$ FOV$^{-1}$ and $n_{swp}v_{swp}$ is in cm$^{-2}$ s$^{-1}$.

The scatter in the $n_{swp}v_{swp}$-LTE relation is not unexpected. The measured scatter in the $n_{swp}v_{swp}$-$\overline{Q_M}$ relation is roughly 40\%. The difference between the reference scan path and a ``real'' scan path is roughly 10\%. Assuming that these variations are independent, that suggests a roughly 50\% variation. The blue lines in Figure~\ref{fig:c_nv} show this variation. It encompasses the bulk of the data points, suggesting that there is no significant extra source of variation in the $n_{swp}v_{swp}$-LTE relation.

We further note the lack of a significant intercept in the $n_{swp}v_{swp}$-LTE relation. This result is actually rather unexpected. The measurement of the LTE rate was done without reference to the solar wind flux as that relation was unknown at the time. One might expect the LTE rate to be systematically underestimated or over-estimated and indeed, there are a number of bright streaks in the LTE-subtracted RASS that suggest that the LTE was underestimated in this process. The lack of an intercept in the $n_{swp}v_{swp}$-LTE relation suggests that the estimation of the LTE, while imperfect for some scans, was not systematically low or high.

Therefore, we find that
\begin{equation} 
C_{R12} \propto  (6.02 \pm 0.31)\times10^{-20} \mathrm{(count~FOV^{-1}~cm^{4})}Q
\label{eqn:nv_lte}
\end{equation}
where $C_{R12}$ is the \rosat\ \oqkev\ band (or R12 band) count rate in count s$^{-1}$ FOV$^{-1}$ and $Q$ has its usual units of cm$^{-4}$ s$^{-1}$. The vignetting of the \rosat\ PSPC produces an effective area for the FOV of 1.56 square degrees \citep{sea2015}. (This value can be found directly from the \rosat\ detector map for the R12 band.) Therefore we find that
\begin{equation}
\varsigma=(3.86 \pm 0.20)\times10^{-20} \mathrm{count~degree^{-2}~cm^{4}}
\end{equation}

\begin{figure}
\includegraphics[width=8.0cm,angle=0.0]{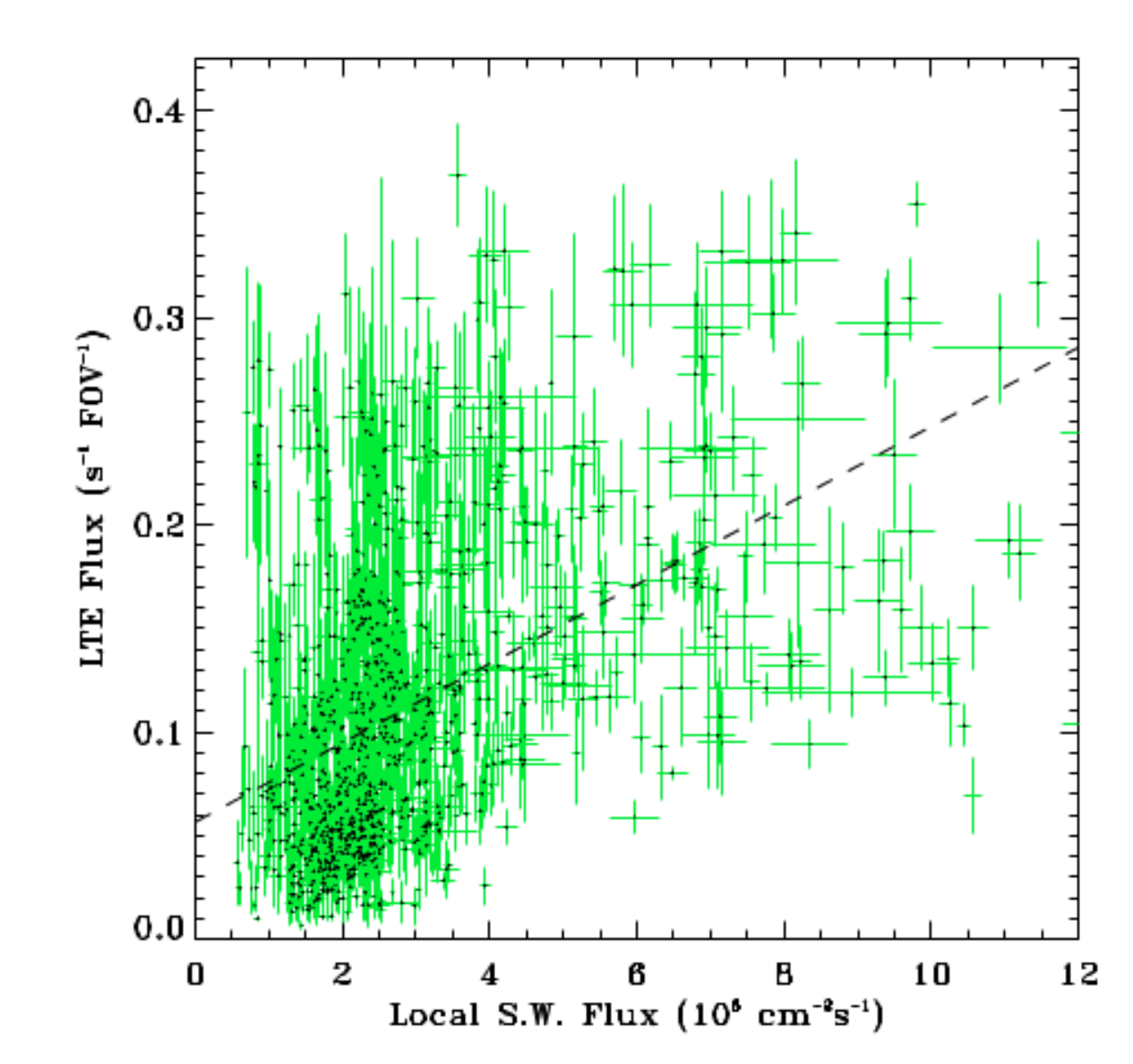}
\caption{The \tqkev\ LTE flux as a function of the local solar wind flux. The {\it dashed} line is the best fit. This fit was used for scaling the solar wind flux to the LTE data in Figure~\ref{fig:m_lc}.
\label{fig:m_nv}}
\end{figure}

The situation for the \tqkev\ band is rather different. Figure~\ref{fig:m_nv} shows the correlation of the \tqkev\ LTE flux and the local solar wind flux. Although there are more low LTE flux data points at low local solar wind values, there is little correlation between the two values, and the Pearson correlation coefficient is only 0.45. In comparison, the Pearson correlation coefficient between the \oqkev\ LTE flux and the solar wind flux is 0.74. The Pearson correlation coefficient between the \tqkev\ and \oqkev\ band LTEs is only 0.14, so the \tqkev\ band LTE emission is better correlated with the local solar wind flux than it is with the LTE emission in the \oqkev\ band. The overall LTE flux in the \tqkev\ band is a factor of seven to eight lower than in the \oqkev\ band, so the uncertainties must consequently be larger. However, we note that the RMS/mean for the \tqkev\ band is similar to that of the \oqkev\ band (33.06 {\it versus} 34.67) so the decrease in the correlation coefficient is not due to an increase in an intrinsic scatter in the \tqkev\ band. The lack of correlation in the \tqkev\ band, compared to the \oqkev\ band, suggests that the oxygen lines that dominate the \tqkev\ are less well correlated with the solar wind flux than the aggregate of the many lines that produce the \oqkev\ band. This is not unreasonable since the oxygen lines will be sensitive to the variation in the oxygen abundance and ionization fraction in the solar wind, while the aggregate of lines in the \oqkev\ lines should be less sensitive. 

\subsection{The Q$_H$ - Solar Wind Flux Relation\label{sec:enlil}} 

\begin{figure*}
\center{\includegraphics[width=18.0cm,angle=0.0]{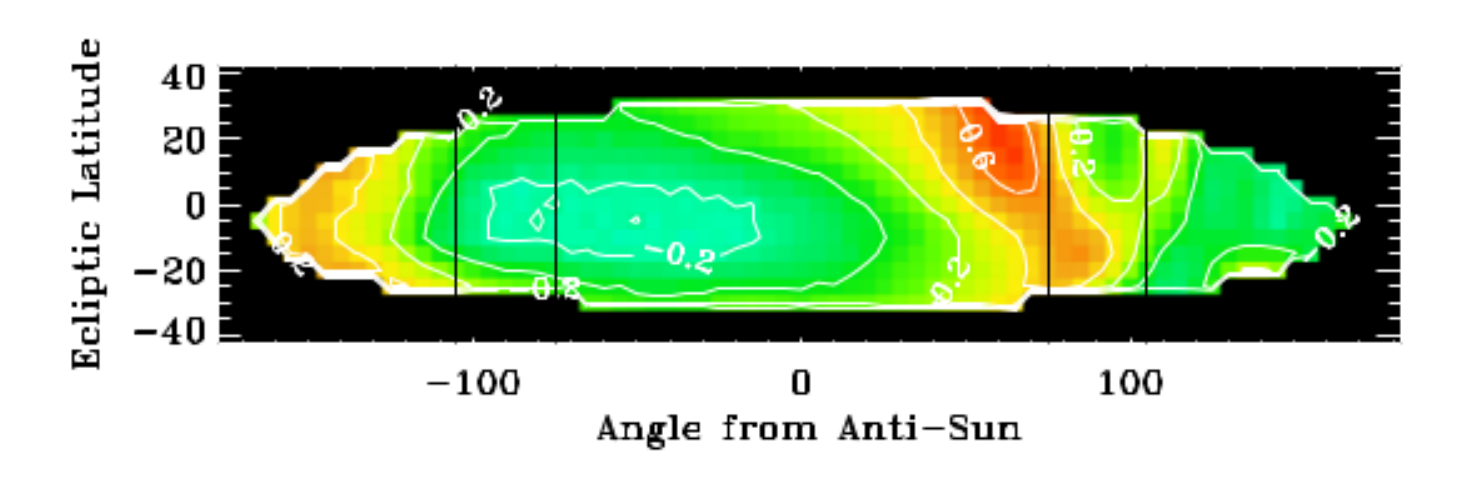}}
\caption{The correlation of the heliospheric SWCX flux and the local solar wind flux as a function of look direction. The longitudinal coordinate is the angle from the anti-Sun while the latitude is the ecliptic latitude. The color scale and the contours are the correlation coefficient between the local solar wind flux (as measured at the Earth in the ENLIL model) and $(n_H+\mathcal{F}n_{He})n_{swp}v_{rel}$ integrated along the line of sight. The two sets of vertical lines mark the locations that can be observed with \rosat\, $\pm15\arcdeg$ from the perpendicular to the Sun; \xmm\ is restricted only to $\pm20\arcdeg$ from the perpendicular to the Sun. The effect of the Parker spiral is apparent at 50\arcdeg to 70\arcdeg as well as between -130\arcdeg and -150\arcdeg. This map is the mean of four maps made from the same ENLIL simulation at $90\arcdeg$ intervals around the Earth's orbit. Each individual map is very similar to the mean shown here.
\label{fig:corr_map}}
\end{figure*}

Combining the $n_{swp}v_{rel}$-LTE and $n_{swp}v_{rel}$-$\overline{Q}$ relation to get the production factor is only valid if the only contribution to the correlation between LTE emission and the local solar wind is the magnetosheath emission. So long as it is not correlated with the local solar wind flux, the heliospheric SWCX will introduce only noise to the relation. Thus, we must investigate the possible correlation of the heliospheric SWCX with the local solar wind flux.

The heliospheric emission along the line of sight is $Q_{H}=\int (n_{ipH}+\mathcal{F}n_{ipHe})n_{swp}v_{rel} dl$, where $n_{ipH}$ is the interplanetary H density, $n_{ipHe}$ is the interplanetary He density, and $\mathcal{F}$ is the ratio of the He charge-exchange cross-section to the H charge-exchange cross-section. Since He cross-sections are usually only a factor of a few smaller than those for H \citep[see the compilation in ][]{dk2006}, we have set $\mathcal{F}=0.5$. The solar wind proton density was extracted from the ENLIL models while the density of the interplanetary neutrals was taken from the model of \citet{dk2006}. Since the ENLIL model was calculated only to 10 au from the Sun within $30\arcdeg$ of the solar equator, we can calculate a line of sight from Earth to 9 au for only a limited number of directions. The distance of 9 au is sufficient for determining the correlation between the heliospheric SWCX and the local solar wind flux: if we assume a heliopause distance of 100 au, then 70\% of the total observed heliospheric emission occurs within the first 9 au. There is some variation between the upwind and downwind directions, but we have calculated this as an average over angle from the upwind direction. 
 
For the region for which we can measure the heliospheric properties over lines of sight 9 au long, we have calculated $Q_{H}$ at $5\arcdeg$ spacing in both ecliptic latitude and ecliptic longitude for each ENLIL time step. For each line of sight we calculated the Pearson correlation coefficient between the $Q_{H}$ and the local $n_{swp}v_{swp}$, as recorded in the model at the location of the Earth. Figure~\ref{fig:corr_map} shows the correlation coefficient as a function of ecliptic latitude and angle from the Sun along the ecliptic. The pattern shown in this map is reasonably independent of the details of the recent solar wind history and the time of year.

The strongest correlations are found between $100\arcdeg$ and $130\arcdeg$ from the Sun, measuring in a positive sense around the ecliptic, or 50\arcdeg to 80\arcdeg from the anti-sun as seen in Figure~\ref{fig:corr_map}. This is the region where one is looking along the Parker spiral beyond the Earth's orbit. Similarly, angles further than $-130\arcdeg$ from the anti-sun have high correlations because one is looking down the Parker spiral within the Earth's orbit. The curvature of the spiral is stronger within the Earth's orbit than without, so there is a shorter distance at which the line of sight is tangent to the spiral. Only a small part of the Parker spiral region is accessible to either \rosat\ or \xmm , so one should not expect a strong correlation between the local solar wind and the heliospheric SWCX. This statement excludes, of course, strong CMEs.

\begin{figure}
\center{\includegraphics[width=7.0cm,angle=0.0]{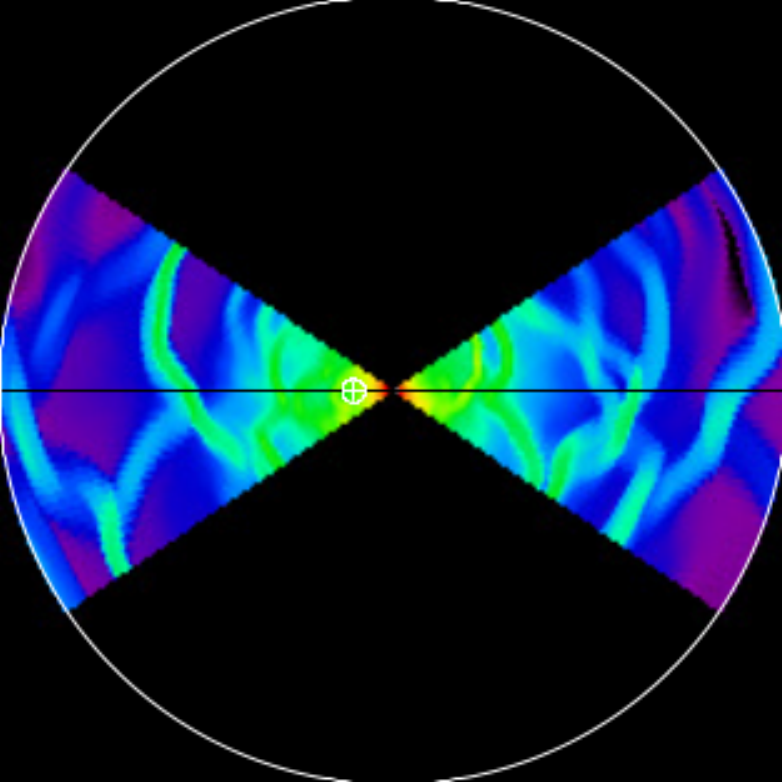}}
\caption{The relative solar wind proton density from a vertical cut through a time step of the ENLIL simulation. The circle has a radius of 10 au. The location of the Earth is marked. 
\label{fig:plane_demo_side}}
\end{figure}

The restriction to within 30\arcdeg of the solar equator is not of great concern for the correlation analysis. As can be seen from the diagrams in \citet{mccomas2003}, during solar minimum, the solar wind between the equator and $\sim30\arcdeg$ is characterized by a dense, slow ``equatorial'' flow, while at higher latitudes, the there is a faster, more tenuous ``polar'' flow. As can be seen in Figure~\ref{fig:plane_demo_side} there are coherent structures extending out of the plane. However, unlike the Parker spiral, these structures have a wide range of tilts, so the direction in which one looks along the structure is not fixed. Further, for a line of sight towards the poles, the equatorial flow accounts for only $\sim30$\% of the heliospheric SWCX flux, assuming an emissivity that declines as $r^{-2}$ as measured from the Sun. Thus, in general, the correlation between the local solar wind and the SWCX observed towards the ecliptic poles should be weak, though SWCX emission will be correlated with the local solar wind for some limited directions at some times. During the solar maximum, which is more relevant for the RASS observations, the differentiation between equatorial and polar flows is missing, and the solar wind speeds and densities are strongly variable in both regions. In this case we again do not expect correlation between the locally measured solar wind and the solar wind at the higher solar latitudes not covered by the ENLIL model.

\section{Discussion}

\subsection{On the production factor}

In \S3.1 and \S3.2 we fitted the $n_{swp}v_{swp}$-LTE relation and the $n_{swp}v_{swp}$-$\overline{Q_M}$ relation, and used the slopes of those relations to determine the \rosat\ \oqkev\ count rate as a function of $Q_M$, and thus the \oqkev\ production factor. Further, we have shown that the scatter in the $n_{swp}v_{swp}$-LTE relation is comparable to that in the $n_{swp}v_{swp}$-$\overline{Q_M}$ relation.

\citet{ir2009} calculated \rosat\ production factors for a ``slow'' solar wind from the available atomic data. The slow solar wind conditions \citep[see][for definition]{vons2010} are appropriate for the equatorial flow at solar minimum. Although the solar wind experienced by the Earth at solar maximum is a complex mixture of slow and fast solar wind, the distribution of velocities during the time of the RASS is consistent with the slow wind, and it is common practice to approximate the equatorial solar wind, even at solar maximum, with the slow solar wind ion/abundance ratios. The calculated \oqkev\ production factor from \citet{ir2009} for charge-exchange with H is $23.64\times10^{-25}$ count  arcmin$^{-2}$ cm$^4$, ($0.851\times10^{-20}$ count degree$^{-2}$ cm$^4$) which is a factor of 4.5 lower than that derived here. 

This disagreement is not altogether surprising given the uncertainty in the atomic data.  A comparison, for example, of the soft X-ray SWCX spectrum shown in \citet{ir2009} with that of \citet{klrk2009} shows strong differences despite a reliance on similar atomic data. A conversion of the \citet{klrk2009} spectrum into a production factor yields $2.13 \times10^{-20}$ count degree$^{-2}$ cm$^4$, which is a factor of only 1.79 lower than that empirically obtained from the LTE data. Addition of further lines to the theoretical spectrum will decrease this difference.

We do not find any combination of systematic uncertainties that could push the measured production factor, $\varsigma$, significantly lower. Had there been a heliospheric contribution to the $n_{swp}v_{swp}$-LTE relation, then the derived $\varsigma$ would have been an upper limit. However, as shown in \S3.3, the heliospheric emission does not contribute to the $n_{swp}v_{swp}$-LTE correlation. Conversely, if the plasmaspheric contribution in the \bats\ simulations were not completely removed, then the derived $\varsigma$ would be a lower limit. As can be seen in Figure~\ref{fig:yz_demo}, the region removed for the plasmasphere, while adequate for that purpose, does not extend all the way to the magnetopause. An alternate cleaning which removes both the plasmasphere and the bulk of the low density region interior to the heliopause does not significantly change the calculated $\overline{Q_M}$, though some individual lines of sight changed significantly. Thus, the plasmasphere removal method does not significantly change the derived production factor. Similarly, if the \citet{hodges1994} model overestimates the exospheric density, as suggested by \citet{oea2003}, then the derived $\varsigma$ would again be a lower limit.

The derived production factor is also a lower limit as we have not accounted for the loss of ions due to charge-exchange as they travel through the exosphere. This is, however, a negligible effect. Assuming a path through the magnetosheath in the GSE-XY plane from the sub-solar point around the Earth to roughly $120\arcdeg$ from the sub-solar point (a path-length of $\sim55$ R$_E$) and the O$^{+7}$ charge-exchange cross-section for a slow solar wind from \citet{dk2006} ($\sigma=3.4\times10^{-15}$ cm$^2$), we find that less than a tenth of a percent of the ions recombine, and we have not even corrected this value for the number of O$^{+8}$ ions that will have charge-exchanged to form O$^{+7}$ ions. Thus, correction of the ion density for the effects of charge-exchange is inconsequential.

\subsection{On the relative contributions of the magnetosheath and the heliosphere to SWCX ``contamination''}

\begin{figure}
\center{\includegraphics[width=8.0cm,angle=0.0]{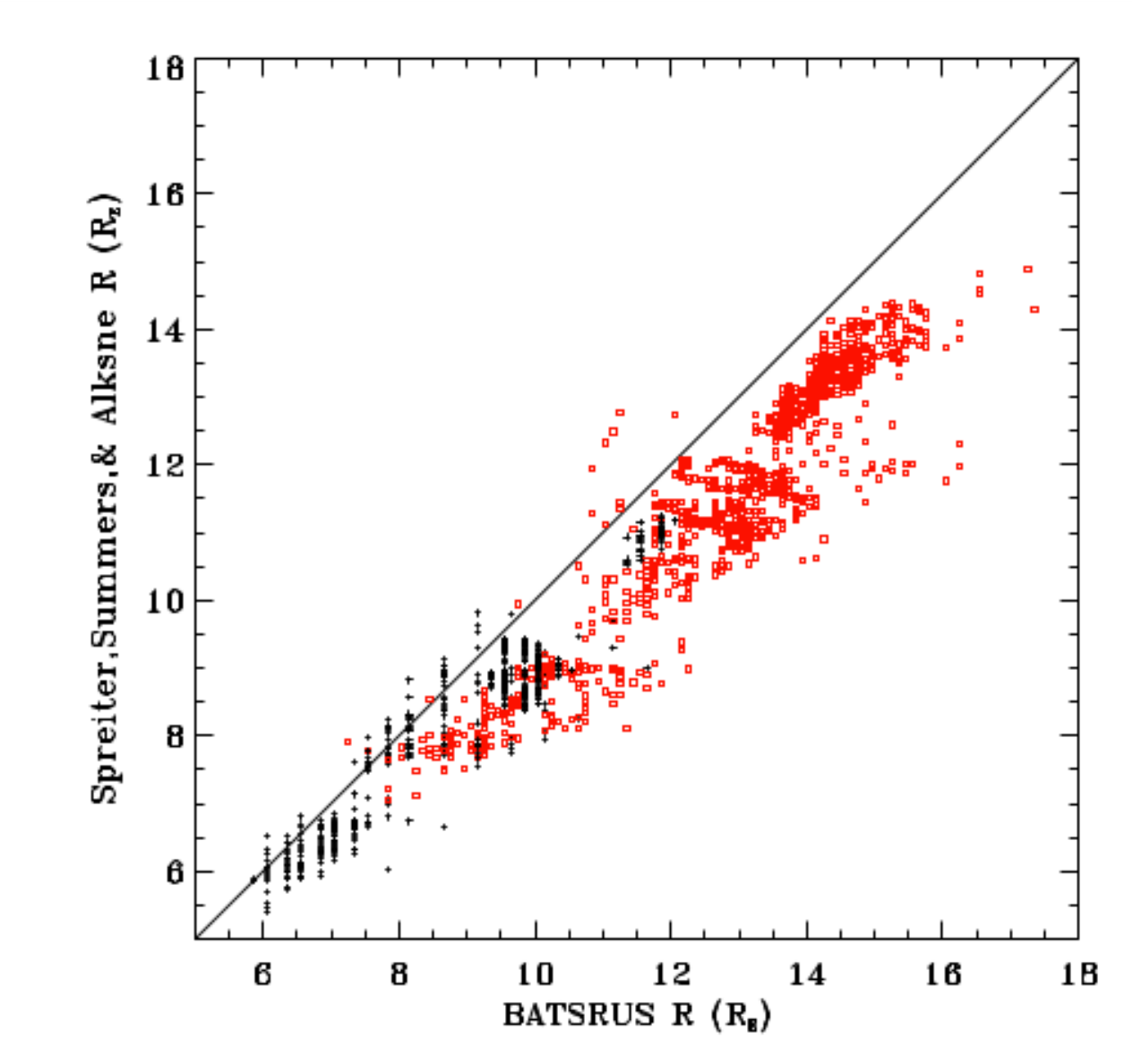}}
\caption{A comparison of the magnetosheath size between the static \citet{ssa1966} model and the \bats\ model. The magnetopause distances are plotted with {\it black crosses} while the bowshock distances are plotted by {\it red boxes}. These were calculated from a subsample of the \bats\ simulations covering typical solar wind conditions. In all cases the distances are measured from the Earth along the GSE-X axis.
\label{fig:spreiter}}
\end{figure}

Efforts to understand the SWCX in the \chandra\/\xmm\ era have been primarily driven by the problem of removing the SWCX emission from observations of diffuse emission. \citet{ks2008} used multiple \xmm\ observations with the same pointing but different epochs and observing geometries to show that SWCX emission is not solely a function of the local solar wind flux; some observations with low local solar wind flux showed SWCX emission while the majority of the observations at those solar wind fluxes did not. Further, the observations that showed SWCX emission when the local solar wind flux was low often had lines of sight that avoided the high emissivity regions of the magnetosheath. However, they found no high local solar wind flux observations ($nv\gtrsim10^9 \flux $) without SWCX emission. Finally, there were a number of observations that should have had lines of sight passing through the highest emissivity regions of the magnetosheath which showed no elevated SWCX emission, while other such observations did. Thus \citet{ks2008} concluded that high local solar wind flux was necessary but not sufficient to predict SWCX emission. Although that SWCX emission observed when the local solar wind was low was probably due to local heliospheric emission, they could not exclude magnetospheric production.

\citet{cs2008} and \citet{csr2011} combed the \xmm\ archive for observations affected by SWCX. They searched for observations where the 0.5-0.7 keV band containing the oxygen lines showed variation not seen in a 2.5-5.0 keV ``continuum'' band. Their results were similar to those of \citet{ks2008}, finding that observations containing SWCX emission tended to occur when \xmm\ was between the Earth and the Sun, but were not limited to such lines of sight. For each observation they also compared the expected magnetosheath flux, as calculated from a static model of the magnetosheath  and the \ace\ solar wind data. They found very poor correlation between the two values, suggesting both magnetosheath and heliospheric contributions. 

Both \citet{ks2008} and \citet{csr2011} used the \citet{ssa1966} model for the size of the magnetosheath given the solar wind density, speed, and temperature. We have compared the magnetopause distance and the bowshock distance of the \citet{ssa1966} model with that found in the \bats\ simulations. The bowshock distance in the \bats\ simulations was set to be the radius along the GSE-X axis at which the pressure drops to less than a tenth of the maximum pressure. The magnetopause distance was set to be the maximum radius along the GSE-X axis at which the magnetic field lines are closed. Figure~\ref{fig:spreiter} compares the \citet{ssa1966} and \bats\ values; the \citet{ssa1966} consistently underestimates the magnetopause and bowshock distances compared to those found by \bats\ . In the case of the bowshock, the underestimate is often 2-3 R$_E$, while the magnetopause is usually off by $\sim1$ R$_E$. Thus, it is likely that both \citet{ks2008} and \citet{csr2011} found a poor correlation between lines of sight passing through the nose of the magnetosheath and SWCX contamination because the lines of sight were actually passing several R$_E$ behind the region of substantial emission. \citet{hs2010,hs2012} extracted ``blank sky'' spectra from all possible \xmm\ observations in order to study the Galactic halo. To study the SWCX emission they considered the oxygen line fluxes for cases of multiple observations of the same target. They looked, in particular, at the difference between the flux of any particular observation and the minimum flux for that direction. This quantity showed little correlation with the $Q$ from the magnetosheath, in part because their model of the magnetosheath \citep{ssa1966} had a stand-off distance fixed to 10 R$_E$ without regard to the solar wind flux; their analysis suffered from a more extreme form of the problem suffered by \citet{ks2008}.

Thus we conclude that many studies seeking SWCX emission from the sub-solar point of the magnetosheath have been looking in the wrong location. Whether the enhanced SWCX emission is observed when looking at the correct location has not yet been determined; a study of SWCX emission in the \xmm\ archive using MHD models is currently ongoing at the University of Leicester.

\citet{hs2010,hs2012} also found ``no universal association between enhanced SWCX emission and increased solar wind flux", a finding consistent with previous studies. Given the lack of correlation between the local solar wind flux and $Q_{H}$ shown in \S3.3, this is not surprising. (Note that \xmm\ observes only a slightly larger region of the sky than was accessible to \rosat\ .) However, since we have shown that the magnetosheath emission produces a strong correlation between the \oqkev\ LTE flux and the local solar wind, but does {\it not} produce as strong a correlation between the \tqkev\ LTE flux (dominated by the emission from the \ion{O}{7} and \ion{O}{8} lines) and the local solar wind flux, then the \tqkev\ LTE flux must be governed by some additional parameter, most likely the oxygen ion abundances in the solar wind. If the ion abundance variation is enough to remove the bulk of the correlation seen from the magnetosheath, then one would expect even less correlation with local heliospheric emission. Unfortunately, the O$^{+8}$ abundance is poorly measured by \ace\ , and the O$^{+7}$ is not much better measured, so it will be difficult to study this issue.

\begin{figure*}[t!]
\center{\includegraphics[width=18.0cm,angle=0.0]{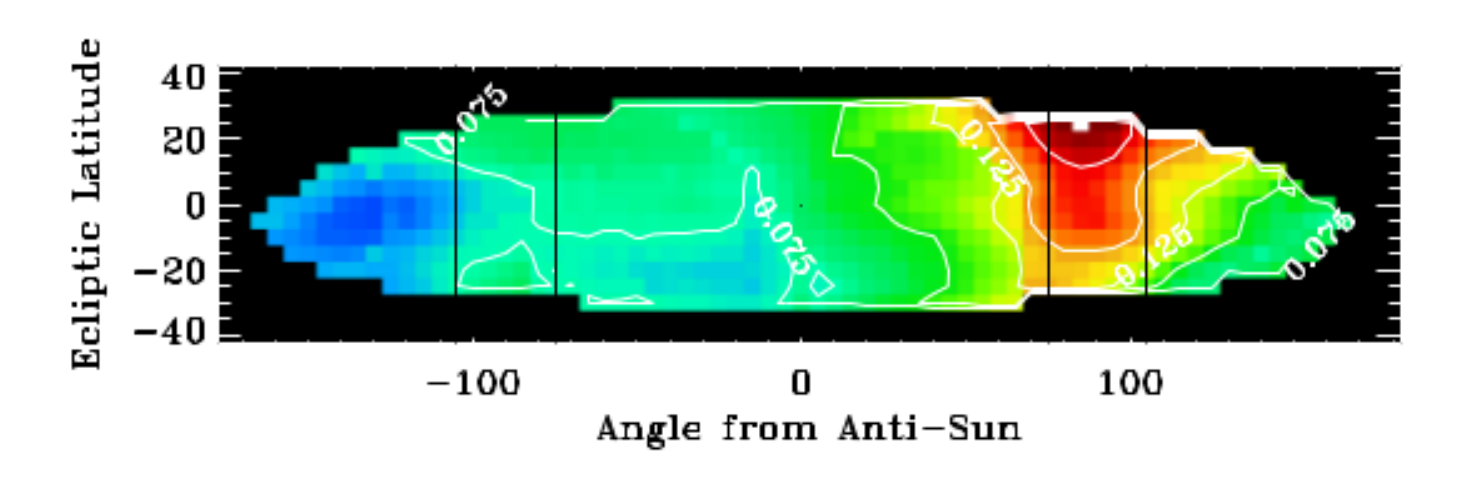}}
\caption{The root mean square of the heliospheric SWCX emission measured over the 559 simulation steps divided by the mean of the heliosperic emission. The coordinate system is the same as in Figure~\ref{fig:corr_map}. This map is the mean of four maps made from the same ENLIL simulation at $90\arcdeg$ intervals around the Earth's orbit. Although the  individual maps are similar to the mean shown here, there is substantial variation. The direction of the He focussing cone has particularly large variation. That signal, which is at a different location in each individual map, was removed before forming this mean map.
\label{fig:vari_map}}
\end{figure*}

In summary, the question of the relative contributions of the magnetosheath and the heliosphere, in general, to SWCX emission events is ill-posed, and studies in the oxygen lines are problematic. There should be a strong correlation between the local solar wind flux and magnetosheath SWCX emission, even if previous attempts to verify that correlation failed due to incorrect expectations for the location of the magnetosheath, and the use of the intrinsically more poorly correlated oxygen lines. Such correlation is not expected, in general, for heliospheric SWCX emission.

\subsection{On SWCX emission prediction and SWCX emission removal}

Since the discovery of the Hubble Deep Field SWCX emission event by \citet{sck2004}, significant effort has been placed on SWCX emission identification in \tqkev\ band observations, either to remove SWCX ``contamination'' for astrophysical reasons, or to identify observations ``blessed'' with SWCX emission for heliospheric reasons. \citet{ks2008} thought it a lost cause to determine whether any arbitrary observation were effected by SWCX without repeated observations of the same field, while \citet{cs2008} showed if an observation was sufficiently long, then SWCX affected intervals could be identified through differential light-curve analysis. \citet{hs2012} applied a local solar wind flux limit to remove SWCX emission affected observations. However, given the correlation analysis for the part of the sky observed by \xmm\ , and excepting CME issues, this limit is as likely to exclude data without significant heliospheric SWCX emission enhancements as it is to include data with significant heliospheric SWCX emission enhancements.

A second method to deal with SWCX emission is {\it ex post facto} modeling of the SWCX emission. Efforts to characterize the emission from the LHB have concentrated on repeated observations of nearby X-ray dark clouds to model and remove the SWCX emission \citep[e.g.][]{dk2011,gupta2009}. \citet{dk2012} has shown, in particular, that modeling the SWCX emission for a line of sight falling entirely within the equatorial flow yields reasonably reliable results, while modeling the SWCX emission for a line of sight dominated by the (poorly characterized) polar flow yields rather disappointing results. We note, however, that these results were obtained at solar maximum when the distinction between polar and equatorial flows becomes problematic. At solar minimum, those flows are much more cleanly distinguished and the polar flow should be much more stable and amenable to modeling.

A third method to deal with SWCX emission is to plan observations in order to minimize (or maximize!) its impact. Reference to Figure~\ref{fig:corr_map} suggests that to model the SWCX emission for an object within $\sim30\arcdeg$ of the ecliptic plane, one would want repeated observations of the object when it lies in the direction of the Parker spiral in order to optimize the correlation between the local solar wind flux and the SWCX emission that is actually observed; observations in other directions will have significantly poorer correlations and thus poorer SWCX emission removal. Conversely, one might wish to observe in such a way as to minimize the variation in the heliospheric flux. This would be particularly useful for shadowing observations where the shadow and the background it is absorbing have to be observed separately. Figure~\ref{fig:vari_map} shows a map of the root mean square of the variation in $Q_H$ normalized by the mean of $Q_H$. Although $\sigma_{Q_H}/\overline{Q_H}$ appears to vary more from simulation to simulation than does the correlation, some features are consistent. The highest variation is perpendicular to the Earth-Sun line in the same quadrant as the Parker spiral beyond the Earth. The opposite direction does not have an elevated variation. Thus it should be noted that high variability region is part of the sky accessible to \xmm\ and \chandra\ , so SWCX variation can be reduced by specifying the part of the year in which the observations are to be made.

\subsection{On remote sensing of the magnetosheath}

\begin{figure}
\center{\includegraphics[width=7.5cm,angle=0.0]{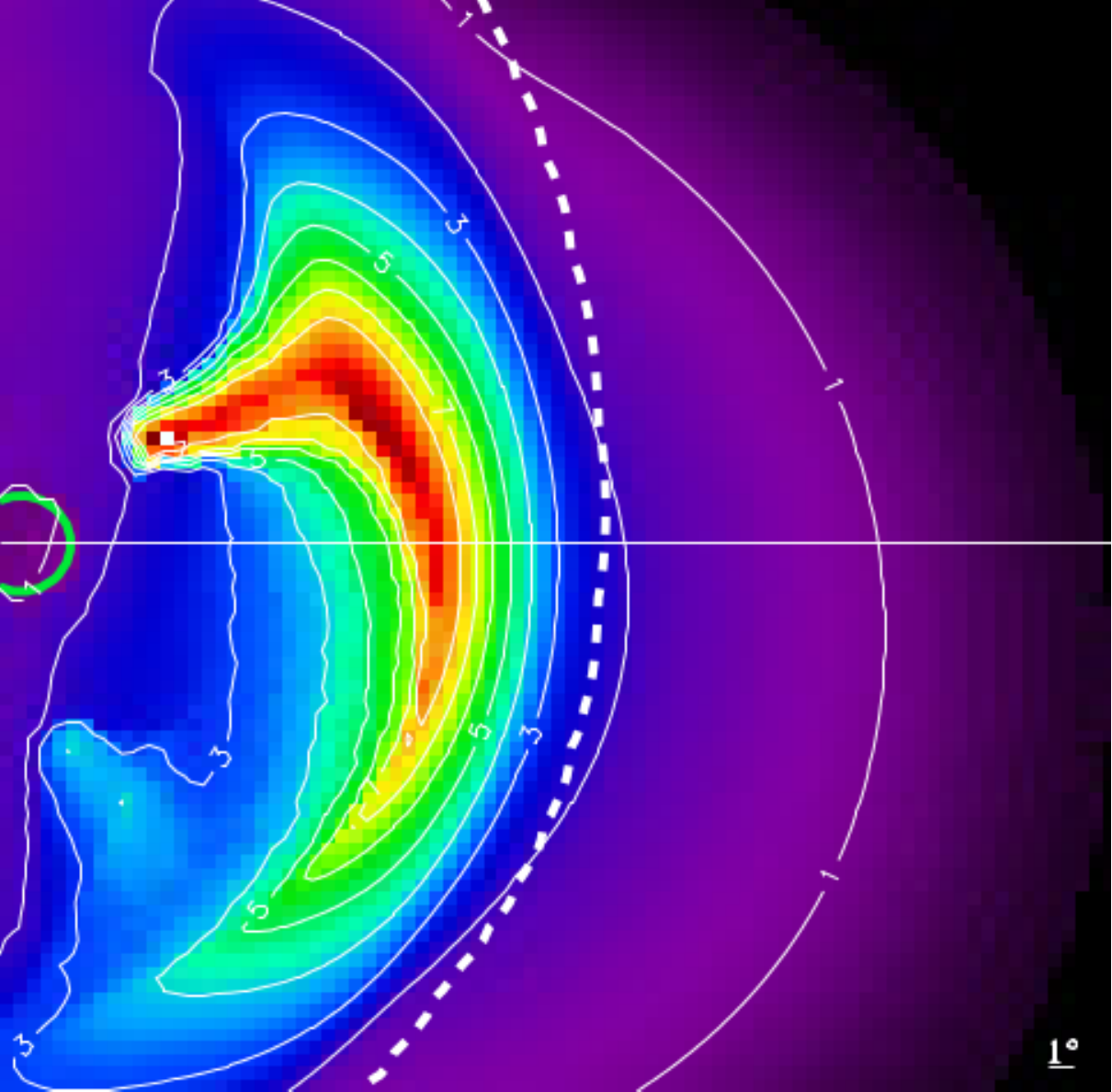}}
\caption{The \oqkev\ flux from the magnetosheath as viewed from GSE coordinate (0,-30,0). The Sun is to the right. The {\it green circle} to the left is the Earth. The {\it white dashed contour} is the projection of the bowshock. The {\it white contours} trace the emission in units of the typical background emission (2.2 count s$^{-1}$ deg$^{-2}$). The {\it straight white line} is the projection of the GSE-X axis. The simulation is for early 2000-07-15 14:25:00, near the solstice, $n_{swp}v_{swp}=1.69\times10^8$ cm$^{-2}$ s$^{-1}$, B$_Z$=-3.97 nT. The simulation was made with $0\fdg5\times0\fdg5$ pixels and is $41\fdg5\times44\arcdeg$. No background has been added to this image.
\label{fig:storm}}
\end{figure}

One of the prime motivations of this work is to pursue a suggestion first made in \citet{csk2005}, that SWCX emission from the magnetosheath could be used for remote sensing of the magnetosheath. This suggestion has spawned a number of mission proposals to observe the magnetosheath, including {\it STORM} \citep{storm2012} and {\it AXIOM} \citep{axiom2012}, as well as smaller proposals to observe the cusps \citep{cuspcube2009}. This is a particularly well posed problem as the SWCX production factor for hydrogen can be derived directly from measurements of the object that we desire to study, albeit in the flanks of the magnetosheath rather than in the much brighter region of interest. 

As shown in Figure~\ref{fig:ms_demo}, the magnetosheath ``breathes'' in and out as the solar wind flux varies. Figure~\ref{fig:storm} demonstrates what would be seen with an X-ray imager 30 R$_E$ from the Earth, along the GSE-Y axis, looking back at the nose of the magnetosheath. The magnetopause and bowshock are clearly defined, as is emission beyond the bowshock due to the exosphere interacting with the free-flowing solar wind. There is a large literature exploring a wide variety of MHD phenomena that should occur within the magnetosheath, such as Kelvin-Helmholtz instabilities and flux-transfer events which may be observable in the X-ray \citep{sun2015}. Whether these structures can be observed depends critically on the count-rate and cadence of such an imager. 

\begin{figure}
\includegraphics[width=9.0cm,angle=0.0]{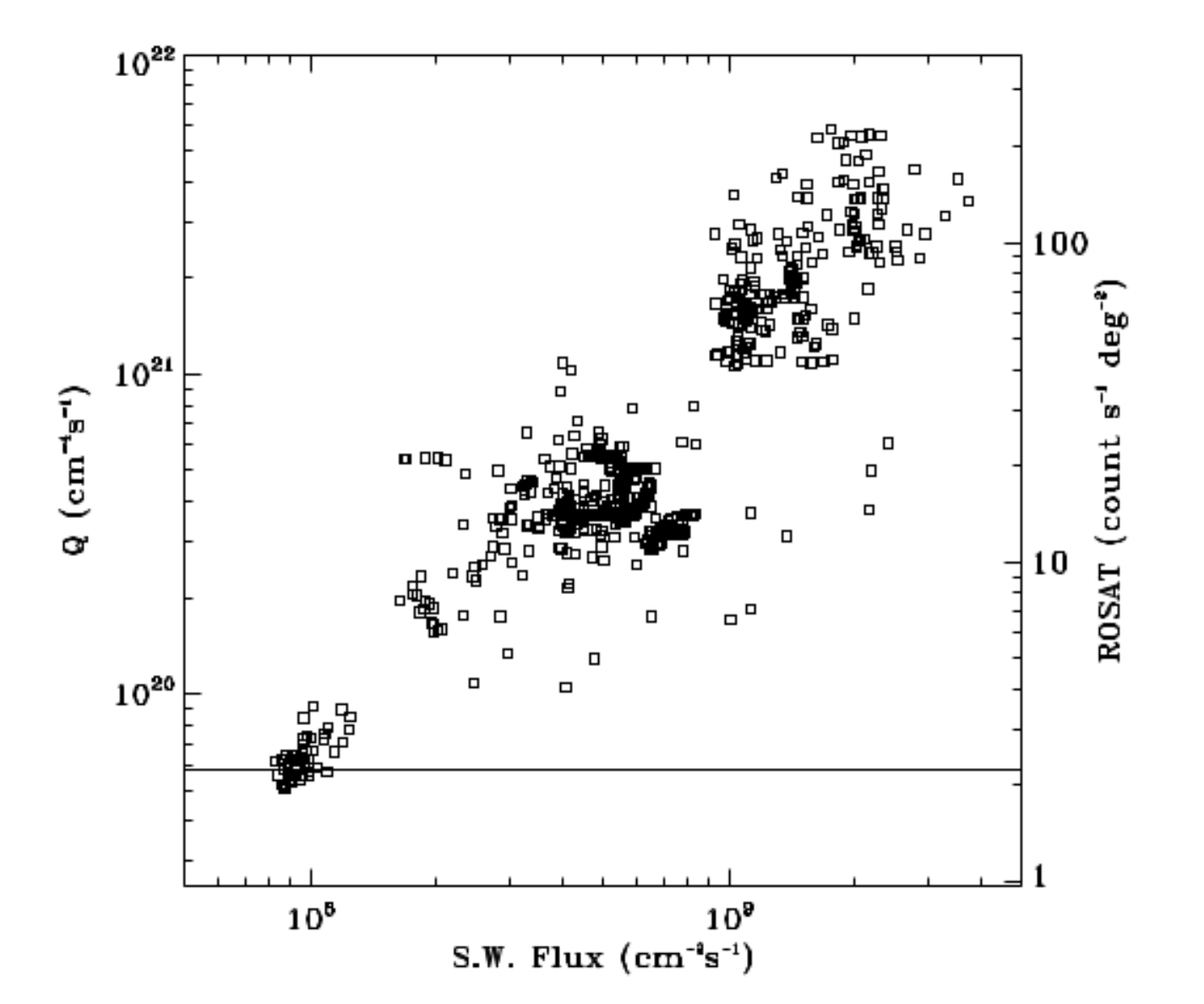}
\caption{The magnetosheath emission from a line of sight through the peak emission on the (projected) GSE-X axis as a function of the solar wind flux. The {\it horizontal line} is the mean \oqkev\ background level.
\label{fig:peak}}
\end{figure}

As a first-order demonstration of the potential utility of such an instrument, we can determine the flux from the magnetosheath and compare it to that of the soft X-ray background. We have calculated the magnetosheath emission as seen from a GSE coordinate (10,-30,0) R$_E$, which produces images such as that seen in Figure~\ref{fig:storm}. For a characteristic value we calculated the emission along a  line of sight through the peak emission on the projected GSE-X axis. On the GSE-X axis the emission at the bowshock is roughly 20\% of that of the GSE-X peak flux. We note that the magnetosheath closer to the dominant cusp tends to be even brighter than the ``subsolar'' flux. We have calculated the characteristic value for a number of simulations and have plotted the result in Figure~\ref{fig:peak}. These simulation values can be compared the background. 

\begin{figure}
\includegraphics[width=8.5cm,angle=0.0]{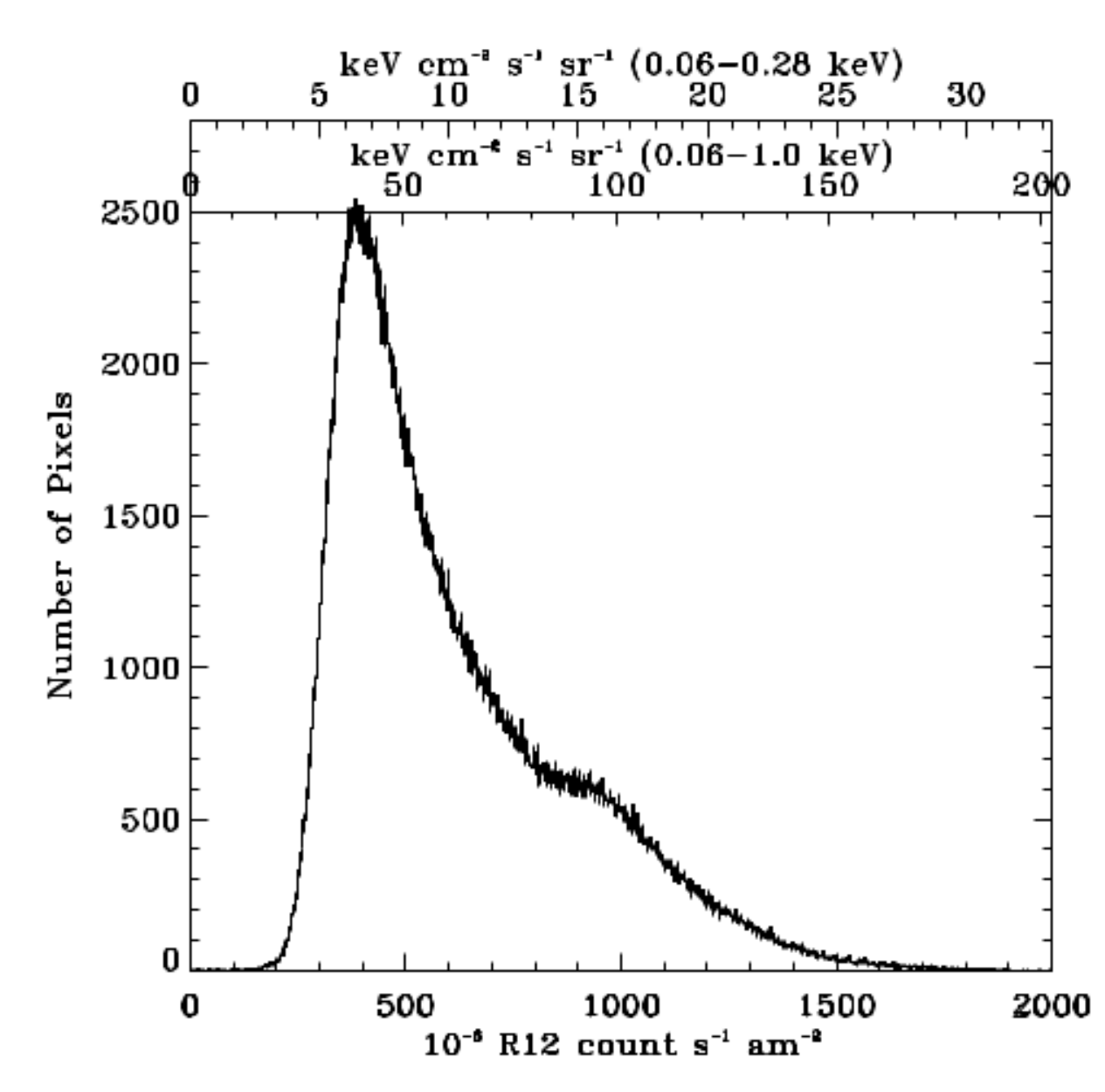}
\caption{The distribution of \oqkev\ surface brightness from the RASS. The conversion from \rosat\ counts to fluxes was made using the spectrum of the soft X-ray background towards the Galactic pole.
\label{fig:rass}}
\end{figure}

As Figure~\ref{fig:rass} shows, the mean \oqkev\ band (R12 band) flux over the entire sky is $6.16\times10^{-4}$ count s$^{-1}$ arcmin$^{-2}$ (2.22 count s$^{-1}$ degree$^{-2}$) while the mode is $3.75\times10^{-4}$ count s$^{-1}$ arcmin$^{-2}$ (1.35 count s$^{-1}$ degree$^{-2}$). These background values are significantly below the ``subsolar'' flux for all but the very calmest solar winds. Thus, imaging the magnetosheath is eminently feasible, given an instrument with a large field of view and modest collecting area.

\section{Summary}

In this work we have calculated the X-ray production factor for solar wind charge-exchange with H. The value is somewhat higher than theoretically calculated values, perhaps because the calculated values did not include all of the faint lines contributing to the emission in the \oqkev\ band. This value is rather robust, being based on the ratio of observed flux to the $Q$ derived from \bats\ simulations; it is {\it particularly} robust for use with other \bats\ simulations being sampled for different observation geometries. As a result, we have shown that remote sensing of the magnetosheath from high Earth orbit, looking back at the magnetosheath, is particularly promising.

On the way to deriving the production factor, we may have resolved a few issues concerning SWCX emission. We have shown how previous models of the magnetosheath SWCX emission are unlikely to agree with \xmm\ observations, simply due to the magnetosheath not being in the expected location. We have also shown why analyses correlating SWCX emission in the \ion{O}{7} and \ion{O}{8} lines with the local solar wind flux have found little correlation; the correlation between the local solar wind and any heliospheric SWCX emission will {\it only} occur in limited portions of the sky {\it and} the oxygen line emission is poorly correlated with the solar wind proton flux due to their sensitivity to abundance and ionization fraction effects. We have also proposed a number of methods to reduce the impact of SWCX emission on various types of observations.

Finally, it must be reiterated that study of SWCX emission in the \tqkev\ band provides little information about the SWCX emission in the \oqkev\ band; the two bands are very poorly correlated because they are produced in very different ways. The \tqkev\ band is due to a small number of lines generated by parent ions with strongly variable abundances. The \oqkev\ band is produced by (currently) innumerable lines from many species. Many of those lines are from L-shell charge exchange for which there is no operational theory \citep{frankel2009}. Until wide-field high spectral resolution imaging becomes common, careful band-averaged studies such as this one will be a far more powerful tool than line studies in the \tqkev\ band for understanding the bulk of the SWCX emission.
 
\acknowledgements

The initial studies that led to this work funded KDK and MRC through National Aeronautics and Space Administration ADAP grant 06-ADP06-32. MRC was partially funded by NASA under grant LSSO06-0032 issued through the Science Mission Directorate's Planetary Science Division. KDK was funded in part through the \xmm\ Guest Observer Facility. KDK would also like to thank the Chemical Heritage Foundation for the hospitality of their reading room for the creation of the first draft of this work.

Part of this work was performed while the primary author and many of the coauthors were attending the International Space Science Institute (ISSI) workshop on solar wind charge-exchange in October 2013 in Bern, Switzerland. Many of the coauthors also attended the ISSI workshop on solar wind charge-exchange in January 2013. 

We are deeply grateful to ISSI for these very productive workshops. Our thanks also to the Community Coordinated Modeling Center at Goddard Space Flight Center for producing the extended ENLIL run, and for providing the large number of \bats\ runs used for this work. The \bats\ and ENLIL simulation results were provided by the CCMC through their public ``Runs on Request'' system (http://ccmc.gsfc.nasa.gov). The CCMC is a multi-agency partnership between NASA, AFMC, AFOSR, AFRL, AFWA, NOAA, NSF and ONR. 

Even with the large number of authors, a study of this type does not occur in a vacuum. We would like to thank all those who, through discussion, argument, and spirited altercations, helped make this a better paper.


\end{document}